\documentclass[10pt, final, onecolumn, twoside,romanappendices]{IEEEtran}
\usepackage{subfig}
\usepackage{ulem}
\usepackage[bookmarks,colorlinks]{hyperref} 
\usepackage{color}
\hypersetup{colorlinks,citecolor= red,filecolor= blue,linkcolor= blue,urlcolor=blue}
\usepackage{graphicx}
\usepackage{tabularx,booktabs}
\newcolumntype{C}{>{\centering\arraybackslash}X} 
\usepackage{algorithm}
\usepackage[noend]{algpseudocode}
\usepackage{dsfont}
\usepackage{amsmath,amssymb}
\usepackage{mathtools}
\usepackage{graphicx}
\usepackage[font=small]{caption}
\usepackage[noadjust]{cite}
\usepackage{amsmath,amsfonts,amssymb,amsthm}
\usepackage{mathtools}
\usepackage{commath}
\usepackage{acronym}  
\acrodef{ofdm}[OFDM]{orthogonal frequency-division multiplexing}
\acrodef{fft}[FFT]{fast fourier transform}
\acrodef{iot}[IoT]{internet of things}
\acrodef{mimo}[MIMO]{multiple-input multiple-output}
\acrodef{siso}[SISO]{single-input single-output}
\acrodef{clt}[CLT]{Central Limit Theorem}
\acrodef{ibi}[IBI]{inter block interference}

\acrodef{ofdma}[OFDMA]{orthogonal frequency-division multiple access}

\acrodef{wmom}[WMOM]{weighted method of moments}

\acrodef{wsom}[WSOM]{weighted second order moment}

\acrodef{pdp}[PDP]{power delay profile}
\acrodef{isi}[ISI]{inter symbol interference}
\acrodef{cp}[CP]{cyclic prefix}
\acrodef{zp}[ZP]{zero padding}
\acrodef{fir}[FIR]{finite impulse response}
\acrodef{v2x}[V2X]{vehicle-to-everything}
\acrodef{nda}[NDA]{non-data-aided}
\acrodef{da}[DA]{data-aided}
\acrodef{ml}[ML]{maximum likelihood}
\acrodef{to}[TO]{timing offset}
\acrodef{wed}[WED]{Weighted Energy Detector}
\acrodef{ed}[ED]{Energy Detector}
\acrodef{rms}[RMS]{root mean square}
\acrodef{tm}[TM]{transition  metric}
\acrodef{ge}[GE]{Gamma estimator}
\acrodef{pdf}[PDF]{probability density function}
\acrodef{cfo}[CFO]{carrier frequency offset}
\acrodef{fom}[FOM]{first-order moment}

\acrodef{iid}[i.i.d]{independent and identically distributed}
\acrodef{mom}[MoM]{Method of Moments} 

\acrodef{som}[SoM]{second order moment} 
\acrodef{sic}[SIC]{successive interference cancelation}
\acrodef{noma}[NOMA]{Non-orthogonal multiple access}
\acrodef{bem}[BEM]{basis expansion model}
\acrodef{ls}[LS]{least squares}
\acrodef{mmse}[MMSE]{minimum mean square error}
\acrodef{pa}[PA]{pilot-aided}
\acrodef{dd}[DD]{decision-directed}
\acrodef{cc}[CE]{channel estimation}
\acrodef{dnn}[DNN]{deep neural network}
\acrodef{mse}[MSE]{mean-squared error}
\acrodef{dl}[DL]{deep learning}
\acrodef{ci}[CI]{channel state information}
\acrodef{mmse}[MMSE]{minimum mean square error}
\acrodef{awgn}[AWGN]{additive white Gaussian noise}
\acrodef{map}[MAP]{maximum a posteriori probability}
\acrodef{ber}[BER]{bit error rate}
\acrodef{kf}[KF]{Kalman filter}
\acrodef{snr}[SNR]{signal-to-noise ratio}
\acrodef{iot}[IoT]{internet of things}

\acrodef{chf}[CHF]{characteristic function}

\usepackage{amssymb}
\usepackage[yyyymmdd,hhmmss]{datetime} 
\newdateformat{monthyeardate}{\monthname[\THEMONTH] \THEDAY, \THEYEAR} 
\usepackage [autostyle, english = american]{csquotes}
\MakeOuterQuote{"}
\newtheorem{theorem}{Theorem}

\usepackage[usestackEOL]{stackengine}
\stackMath
\usepackage{stfloats}
\usepackage{mathtools}
\usepackage{arydshln}
\usepackage{booktabs}
\usepackage{siunitx}

\makeatletter
\let\old@ps@headings\ps@headings
\let\old@ps@IEEEtitlepagestyle\ps@IEEEtitlepagestyle
\def\confheader#1{%
  \def\ps@headings{%
    \old@ps@headings%
    \def\@oddhead{\strut\hfill#1\hfill\strut}%
    \def\@evenhead{\strut\hfill#1\hfill\strut}%
  }%
  \def\ps@IEEEtitlepagestyle{%
    \old@ps@IEEEtitlepagestyle%
    \def\@oddhead{\strut\hfill#1\hfill\strut}%
    \def\@evenhead{\strut\hfill#1\hfill\strut}%
  }%
  \ps@headings%
}
\makeatother

\begin{document}


\title{Draft: Low Complexity Time Synchronization for Zero-padding based Waveforms}
\author{
	\vspace{0.2cm}
   Koosha~Pourtahmasi~Roshandeh,~\IEEEmembership{Student~Member,~IEEE},
   Mostafa~Mohammadkarimi,~\IEEEmembership{Member,~IEEE}, and
    Masoud~Ardakani,~\IEEEmembership{Senior~Member,~IEEE}
}
\maketitle


\begin{abstract}
~The discussion on using \ac{zp} instead
of a \ac{cp} for enhancing channel estimation and
equalization performance is a recurring topic in waveform design for future wireless systems that high spectral efficiency and location awareness are the key factors. 
This is particularly
true for orthogonal signals, such as \ac{ofdm}. \ac{zp}-\ac{ofdm} is appealing for joint communications and sensing (JCS) in 6G networks because it takes the advantage of both \ac{ofdm} and pulse radar. 
In term of communication, \ac{zp}-\ac{ofdm} compared to \ac{cp}-\ac{ofdm}, has higher power efficiency and lower \ac{ber}. However, time synchronization is challenging in \ac{zp}-\ac{ofdm} systems due to the lack of \ac{cp}. In terms of sensing, \ac{zp} facilitates ranging methods, such as time-sum-of-arrival (TSOA).  
In this paper, we propose a moment-based \ac{to} estimator for \ac{mimo} \ac{zp}-\ac{ofdm} system without the need for pilots. 
We then introduce the 

which significantly improves the estimation accuracy of the previous estimator. We show that the proposed method asymptotically reaches the \ac{ml} estimator. 
 Simulation results show very high probability of lock-in for the proposed estimators under various practical scenarios. 
\end{abstract}

\section{Introduction}

\IEEEPARstart{O}{}rthogonal frequency-division multiplexing (OFDM) technique is widely employed in wireless communication systems, mainly due to its ability to convert a frequency-selective fading channel into a group of flat-fading sub-channels \cite{lu2000space}. 
Compared to conventional single-carrier systems, \ac{ofdm} 
offers  increased robustness  against  multipath fading distortions  since    channel equalization  can be easily  performed  in  the  frequency domain  through  a  bank  of  one-tap  multipliers \cite{dai2010positioning}. Moreover, \ac{ofdm} can be efficiently implemented using \ac{fft} \cite{murphy2002low}, which makes it more appealing compared to other multi-carrier modulation techniques such as filter bank multi carrier and generalised frequency division multiplexing. 

 Because of its  advantages, \ac{ofdm} is used in many IEEE standards, such as, IEEE 802.15.3a, IEEE  802.16d/e,  and IEEE 802.15.4g  \cite{green, jimenez2004design, ofdm2ieee802.15}. In addition, \ac{ofdm} combined with massive \ac{mimo} technique achieves a high data rate, making it suitable for multimedia broadcasting \cite{kim2008apparatus}. Moreover, many \ac{iot} applications such as smart buildings and \ac{v2x} leverage \ac{ofdm} as their main communication scheme \cite{ofdmieee802.15, ofdm2ieee802.15 }.
 
 \ac{ofdm}, however, is susceptible to sever \ac{isi} caused by the high selectivity of the fading channel \cite{ wang2005robust}. In order to mitigate this issue, usually a guard interval with a fixed length is inserted between every two consecutive \ac{ofdm} symbols. When the guard interval is the partial repetition of the transmitting data samples, this scheme is called cyclic prefix (\ac{cp})-\ac{ofdm} \cite{channelestimationCP }. When the guard interval is filled with zeros, the scheme is called zero-padded (\ac{zp})-\ac{ofdm} \cite{muquet2000reduced}. 
 
The discussion on using \ac{zp} instead
of a \ac{cp} for enhancing channel estimation and
equalization performance is a recurring topic. This is particularly
true for orthogonal signals, such as \ac{ofdm}. 
 The primary benefit of \ac{cp}-\ac{ofdm} over \ac{zp}-\ac{ofdm} is the ease of timing offset (\ac{to}) estimation or equivalently estimating the starting point of the received samples to apply fast Fourier transform (\ac{fft}). This is referred to as time synchronization \cite{tufvesson1999time}, and is easily carried out by using \ac{cp} and its correlation with the data sequence. Despite the ease of time synchronization, \ac{cp}-\ac{ofdm} has some major disadvantages such as extra power transmission and a higher \ac{ber} compared to  \ac{zp}-\ac{ofdm} \cite{giannazpcp}, which are due to the transmission of \ac{cp}. While \ac{zp}-\ac{ofdm} does not have such drawbacks, its time synchronization or equivalently \ac{to} estimation is very difficult and complicated \cite{koosha2020}.
 
 There are two approaches in order to estimate  \ac{to}  in \ac{zp}-\ac{ofdm}. In the first approach, called \ac{da} time synchronization, a series of training sequences (pilots) are used to estimate \ac{to}. The second approach, referred to as \ac{nda} time synchronization, relies on the statistical properties of the transmitted data sequence. In the next subsection, we briefly review \ac{to} estimation methods proposed for both \ac{da} and \ac{nda} time synchronization in \ac{zp}-\ac{ofdm} systems.

 \subsection{Related work}
 
 The \ac{da} time synchronization for \ac{zp}-\ac{ofdm} has been studied in the literature \cite{nasir2016timing}, where a highly correlated training sequence (pilot) is employed in order to increase the auto-correlation of the received signal, which is then used for estimating the \ac{to}. Such a pilot-based time synchronization algorithm  can achieve reliable performance while having a reasonable complexity \cite{nasir2016timing}. For \ac{nda} time synchronization in \ac{zp}-\ac{ofdm}, however, a low-complexity TO estimation algorithm with an accuracy comparable to that of CP-OFDM counterparts does not exist. Existing \ac{nda} time synchronization algorithms for \ac{zp}-\ac{ofdm}  \cite{bolcskei2001blind, LeNir2010} mainly detect the jumps in the energy of the signal. Such methods are heuristics that detect the point that the energy\footnote{Sometimes two sliding windows are used and the change in the energy ratio of these two windows is tracked.} of the samples in the window drops significantly. However, jump-based techniques greatly suffer from the natural randomness of the received samples; thus, exhibit poor performance in terms of probability of lock-in, i.e. correct \ac{to} estimation. A mathematical approach towards \ac{nda} \ac{to} estimation for \ac{zp}-\ac{ofdm} systems has been proposed in \cite{koosha2020}. The authors in \cite{koosha2020} proposed a \ac{ml} \ac{to} estimator for \ac{zp}-\ac{ofdm} under a frequency selective channel. However,  the algorithm in  \cite{koosha2020} is highly complex which hinders its implementation for \ac{mimo} systems or even single-antenna mobile users. Moreover, the algorithm in  \cite{koosha2020} cannot be used for low \ac{snr} as the proposed expressions in the estimator yield infinity due to floating-point errors.

 \subsection{Motivation}

\ac{zp}-\ac{ofdm}  has several advantages compared to \ac{cp}-\ac{ofdm} \cite{giannazpcp}. For example, regardless of the channel nulls, it is possible to perform finite impulse response equalization in \ac{zp}-\ac{ofdm} systems \cite{giannazpcp}. Moreover,  channel estimation and tracking is easier in \ac{zp}-\ac{ofdm} compared to \ac{cp}-\ac{ofdm} \cite{giannazpcp}. Finally, \ac{zp}-\ac{ofdm} requires less transmission power compared to \ac{cp}-\ac{ofdm}, due to lack of \ac{cp}, which makes it a suitable candidate for power-limited devices. However, time synchronization becomes challenging in \ac{zp}-\ac{ofdm} where proposed \ac{nda} algorithms in the literature fail to achieve a high lock-in probability, or practical  complexity. Hence, an accurate yet low-complexity \ac{nda} time synchronization algorithm for \ac{zp}-\ac{ofdm}   is needed. The goal of this paper is to fill this existing gap.

\subsection{Contributions}

In this paper, we first propose a moment-based \ac{nda} \ac{to} estimator for \ac{mimo} \ac{ofdm}. In general, moment estimators are derived via solving equations involving the theoretical moments of the received samples and their natural moment estimators \cite{kay1993fundamentals}. 
In this paper, we choose $n = 2$ in order to keep the complexity of the estimator very low. Later, we propose a weighted version of the moment estimator to further improve the the probability of lock-in.  The contribution of the paper is summarized as follows
\begin{itemize}

    \item an  \ac{nda} \ac{to} estimator based on method of moments (MOM) for \ac{mimo} \ac{zp}-\ac{ofdm} systems in doubly selective channels is proposed. This algorithm 

    \begin{itemize}
        \item achieves high  lock-in probability, 
        
        \item  has significantly lower complexity compared to \cite{koosha2020}; thus, it can be employed in \ac{mimo} and even massive-\ac{mimo} systems.
        
        \item is suitable for deployment in very low \ac{snr}s in contrast to \cite{koosha2020} and \textcolor{red}{TM},

    \end{itemize}
    
   
    \item a weighted moment-based \ac{nda} \ac{to} estimator for \ac{mimo} \ac{zp}-\ac{ofdm} systems in highly selective channels is proposed. This algorithm 

    \begin{itemize}
        \item has all the benefits of the first estimator while significantly improving its lock-in probability
    \end{itemize}

\end{itemize}{}

This paper is organized as follows. The main ideas, the proposed estimators, and the complexity of the estimators  are presented in Section \ref{sec: ml estimation}. Simulation results and conclusions are given in Sections \ref{sec : simulation results}  and \ref{sec : conclusion}, respectively.

\textit{Notations}: Column vectors are denoted by bold lower case letters. Random variables are indicated by uppercase letters. Matrices are denoted by bold uppercase letters. 
Conjugate, absolute value, transpose, and the expected value are indicated by $(\cdot)^*$, $|\cdot|$, $(\cdot)^{\rm{T}}$, and $\mathbb{E}\{\cdot\}$, respectively. Floor function is denoted by $\lfloor \cdot \rfloor$. Brackets, e.g. ${\bf a}[k]$,  are used for discrete indexing of a vector  ${\bf a}$. Natural estimator of an observation sample is defined as
 
\begin{equation} \nonumber
\mathbb{E} \{ \lvert x[i] \rvert^n \} = \frac{1}{m} \sum_{i=1}^{m} \lvert x[i] \rvert^n.    
\end{equation}



\subsection{System model} \label{sec: sys mod}

We consider a \ac{mimo}-\ac{ofdm} wireless system with $m_{\rm{t}}$ and $m_{\rm{r}}$ transmit and receive antennas, respectively. This system uses \ac{zp}-\ac{ofdm} technique for JCAS over a frequency selective Rayleigh fading channel. 
Let 
$\{x^{(n,k)}_i\}_{k=0}^{n_{\rm{x}}-1}$ with  $\mathbb{E}\{|x^{(n,k)}_i|^2\}= \sigma^2_{\rm{x}}$, denote the  $n_{\rm{x}}$ complex data samples  from the $n$-th \ac{ofdm} block to be transmitted  from the $i$-th transmit antenna.  The corresponding \ac{ofdm} signal can be expressed as 
\begin{align}\label{eq: ofdm symbol}
x^{(n)}_i(t)=\sum_{k=0}^{n_{{\rm{x}}-1}} x^{(n,k)}_i e^{\frac{j2\pi k t}{T_{\rm{x}}}}\,\,\,\,\,\,\,\,\,\,\ 0 \le t \le T_{\rm{x}},
\end{align}
\noindent where  $T_{\rm{x}}$ denotes the duration of the data signal. By adding a zero-padding guard interval of length $T_{{\rm{z}}}$ to \eqref{eq: ofdm symbol}, the $n$-th transmitted \ac{zp}-\ac{ofdm} block from the $i$-th transmit antenna is given as 
\begin{align} \label{eq: s continues}
s^{(n)}_i(t)=
\begin{cases}
x^{(n)}_i(t)  \,\,\,\,\,\,\,\,\,\,\,\ 0 \le t \le T_{\rm{x}}  \\
0 \,\,\,\,\,\,\,\,\,\,\,\,\,\,\,\,\,\,\,\,\,\,\,\,\,\,\,   T_{\rm{x}} < t \le T_{\rm{s}},
\end{cases}
\end{align}
\noindent where $T_{\rm{s}}$ denote the symbol duration, and  $T_{\rm{s}}= T_{\rm{x}}+T_{\rm{z}}$.


Let  $f_{\rm{s}}=1/T_{\rm{sa}}$ denote the sampling rate at the receiver.  In the absence of synchronization error,  the discrete received baseband vector of the $n$-th OFDM block is expressed as \cite{wang2022channel}
\begin{align}\label{Sys Model: matrix form conv 2}
{\bf y}^{(n)}=
\begin{cases}
{\bf H} {\bf s}^{(n)} + {\bf w}^{(n)},  \,\,\,\,\,\,\,\,\,\,\,\ n \ge 0  \\
{\bf w}^{(n)}, \,\,\,\,\,\,\,\,\,\,\,\,\,\,\,\,\,\,\,\,\,\,\,\,\,\,\,\,\,\,\,\,\,\,\,\,\,\,  n<0,
\end{cases}
\end{align}
where 
${\bf H}$ denotes the discrete channel matrix, and is defined as 
\begin{equation} \label{matrix H}
 {\bf H}=\begin{pmatrix}\vspace{0.2 cm} 
{\bf H}_{11} & {\bf H}_{12} & \cdots & {\bf H}_{1m_{\rm{t}}}  \\ \vspace{0.2 cm}
{\bf H}_{21} & {\bf H}_{22} & \cdots & {\bf H}_{2m_{\rm{t}}}  \\ \vspace{0.2 cm}
\vdots & \cdots & \ddots & \cdots  \\ \vspace{0.2 cm}
{\bf H}_{m_{\rm{r}} 1} & {\bf H}_{m_{\rm{r}} 2} & \cdots & {\bf H}_{ m_{\rm{r}} {m_{\rm{t}}}}  \\
\end{pmatrix}, 
\end{equation}
\noindent 
where ${\bf H}_{ji}$ is the $n_{\rm{s}}\times n_{\rm{s}}$ lower triangular Toeplitz channel matrix between the transmit antenna $i$ and the received antenna {$j$, with first column  $[h_{ji}[k,0] \ h_{ji}[k,1] \ \cdots \  h_{ji}[k,n_{\rm{h}}-1] ~ \  0 \  \cdots \ 0]^\text{T}$ where $0\le k \le n_{\rm{s}}-1$, and $\{h_{ji}[k, l]\}_{l=0}^{n_{\rm{h}}-1}$ denote  the $n_{\rm{h}}$ channel taps  between the transmit antenna $i$ and the received antenna $j$ at time $k$ and delay $l$ \cite{wen2022ergodic}. 
The channel is considered to be
Rayleigh fading, and 
the channel taps are assumed to be statistically independent and are modeled by zero-mean complex Gaussian random variables with the autocorrelation function 
\begin{align}\label{7u8i0000}
&\mathbb{E}{\big{\{}}h_{j_1i_1}[k_1,l]h_{j_2i_2}^*[k_2,l-m]\big{\}} \\ \nonumber 
\vspace{-0.5em}
&=\sigma_{{{\rm{h}}_l}}^2R[k1-k2]\delta[m]\delta[j_1-j_2]\delta[i_1-i_2],
\end{align}
$l=0,1,\dots, n_{\rm{h}}-1$, $0\le k_1,k_2 \le n_{\rm{s}}-1$, $1\le i_1,i_2 \le m_{\rm t}$, and $1\le j_1,j_2 \le m_{\rm r}$.
In \eqref{7u8i0000}, $R[k1-k2]$ is an arbitrary function, where 
as the relative speed of the transmitter and receiver increases,   $R[k_1-k_2]$ approaches $\delta[k_1-k_2]$. 
The power delay profile of the channel, i.e. $\sigma_{{{\rm{h}}_0}}^2, \sigma_{{{\rm{h}}_1}}^2 \ldots \sigma_{{{\rm{h}}_{n_{\rm h}-1}}}^2$, is assumed to be known at the receiver.
The vectors  ${\bf s}^{(n)}$, ${\bf y}^{(n)}$, and ${\bf w}^{(n)}$ in \eqref{Sys Model: matrix form conv 2} are given as 
\begin{equation} \label{sys mod:  s y w}
{\bf s}^{(n)} =\begin{pmatrix}\vspace{0.2 cm}
{\bf s}^{(n)}_1   \\ \vspace{0.2 cm}
{\bf s}^{(n)}_2   \\ \vspace{0.2 cm}
\vdots   \\ \vspace{0.2 cm}
{\bf s}^{(n)}_{m_{\rm{t}}}   \\
\end{pmatrix},~
{\bf y}^{(n)} \triangleq \begin{pmatrix}\vspace{0.2 cm}
{\bf y}^{(n)}_1   \\ \vspace{0.2 cm}
{\bf y}^{(n)}_2   \\ \vspace{0.2 cm}
\vdots   \\ \vspace{0.2 cm}
{\bf y}^{(n)}_{m_{\rm{r}}}   \\
\end{pmatrix},~
{\bf w}^{(n)} \triangleq \begin{pmatrix}\vspace{0.2 cm}
{\bf w}^{(n)}_1   \\ \vspace{0.2 cm}
{\bf w}^{(n)}_2   \\ \vspace{0.2 cm}
\vdots   \\ \vspace{0.2 cm}
{\bf w}^{(n)}_{m_{\rm{r}}}   \\
\end{pmatrix},
\end{equation}}
where ${\bf s}^{(n)}_i$, ${\bf y}^{(n)}_j$ and ${\bf w}^{(n)}_j$  denote the $n$-th  transmitted ZP-OFDM block  from the $i$-th transmit antenna, the corresponding received vector at the $j$-th receive antenna, and the noise vector at the $j$-th receive antenna, respectively, and are defined as 
\begin{subequations}
\begin{align}
 {\bf y}^{(n)}_j &\triangleq \big{[} y^{(n)}_j[0] \  y^{(n)}_j[1] \ \ldots \ 
 y^{(n)}_j[n_{\rm s}-1] \big{]}^{\rm T},
 \\ 
 \vspace{-2em}
{\bf w}^{(n)}_j & \triangleq \big{[} w^{(n)}_j[0] \  w^{(n)}_j[1] \ \ldots \ 
 w^{(n)}_j[n_{\rm s}-1] \big{]}^{\rm T}
\\
 {\bf s}^{(n)}_i & \triangleq \big{[} s^{(n)}_i[0] \  s^{(n)}_i[1] \ \ldots \ 
 s^{(n)}_i[n_{\rm s}-1] \big{]}^{\rm T},
 \\ \nonumber 
&= [x^{(n)}_i(0)  \
x^{(n)}_i(T_{\rm{sa}})   \
\cdots   \ 
x^{(n)}_i((n_{\rm{x}}-1) T_{\rm{sa}})   \ {\bf 0}_{n_{\rm z}}^{\rm T}]^{\rm T},
\end{align}
\end{subequations}
where $n_{\rm{s}}\triangleq \lfloor T_{\rm{s}}/T_{\rm{sa}} \rfloor$, $n_{\rm{x}}\triangleq \lfloor T_{\rm{x}}/T_{\rm{sa}} \rfloor$, and $n_{\rm{z}}\triangleq \lfloor T_{\rm{z}}/T_{\rm{sa}} \rfloor$ denote the total number samples, number of data samples, and the number of zero-padded samples  per \ac{ofdm} block, respectively, and $n_{\rm{s}} = n_{\rm{x}}+n_{\rm{z}}$. 

In order to avoid \ac{isi}, the length of the zero-padding should be greater than or equal to the number of channel taps, i.e. $n_{\rm{z}} \ge n_{\rm{h}}$. This assumption holds throughout our analysis in this paper. Noise samples are assumed to follow zero-mean complex Gaussian random variable with correlation as 
\begin{equation}
    \mathbb{E}\{w^{(m)}_{j}[k]*w^{(m')}_{j'}[k']\}= \sigma_{\rm n}^2 \delta[m-m']\delta[j-j']\delta [k-k'],
\end{equation}
where 
According to the \ac{clt} \cite{kwak2017central}, the transmitted \ac{ofdm} samples, i.e. $s^{(n)}_i[k]= x^{(n)}_i(kT_{{\rm{sa}}}),~ \forall k \in \{0, 1, \cdots , {n_{\rm{x}}}-1 \}$,  can be modeled as \ac{iid} zero-mean complex Gaussian random variables. Hence,
\begin{align} \label{eq: ofdm samples gauss}
s^{(n)}_i[k] ~\text{or}~ x^{(n)}_i(kT_{{\rm{sa}}}) \sim \mathcal{CN}(0,\sigma^2_{\rm x}),~~~ \forall k \in \{0, 1, \cdots , {n_{\rm{x}}}-1 \}
\end{align}
where 
\begin{align} \label{eq: ofdm samples gauss power}
\mathbb{E}\Big{\{}s^{(n)}_i[k] s^{(n)}_p[k']^*\Big{\}}&=\mathbb{E}\Big{\{}x^{(n)}_i(kT_{{\rm{sa}}}) x^{(n)}_p(k'T_{{\rm{sa}}})^*\Big{\}} \\ 
&= \sigma^2_{\rm x} \delta[k-k'] \delta[i-p],\\ \nonumber &\forall i,p \in \{1,2,\cdots,{m_{\rm{t}}} \}, \\  \nonumber
&\forall k,k' \in \{0, 1, \cdots , {n_{\rm{x}}}-1 \}.
\end{align}
Now, assume that  there is a \ac{to} $\tau \triangleq dT_{\rm{sa}}+\epsilon$ between the transmitter and the receiver, where  $d$ and $\epsilon$ represent the integer and fractional part of the \ac{to}, respectively. Since the fractional part of \ac{to}, $\epsilon$,  can  be  corrected  through channel equalization and carrier frequency offset estimation \cite{morelli2007synchronization}, it suffices to estimate the  integer part of \ac{to}. In fact, it  is  common in practice to model the \ac{to} as a multiple of  the  sampling   period,  and consider the remaining fractional error as part of the channel impulse response. To this end, we focus on estimating the integer part of the \ac{to},  $d$, which is essential in order to perform the \ac{fft} operation at the receiver, and decode the data in subsequent steps.

\ac{TOA Estimation}
In this section, 
The next section proposes the moment-based  estimators for estimating $d$. In order to simplify the notations and derivations, we first consider a \ac{siso} system in the next section, and then extend our discussions to  \ac{mimo} systems.

\section{ Moment-based estimator for \ac{siso}-\ac{ofdm}} \label{sec: siso}
For the SISO system, the received signal in \eqref{Sys Model: matrix form conv 2} can be rewritten as  
\begin{align}\label{eq : siso matrix form conv 2}
{\bf y}^{(n)}=
\begin{cases}
 {\bf{H}} {\bf s}^{(n)} + {\bf w}^{(n)} \triangleq  {\bf v}^{(n)} +{\bf w}^{(n)},  \,\,\,\,\,\,\,\,\,\,\,\ n \ge 0  \\
{\bf w}^{(n)}, \,\,\,\,\,\,\,\,\,\,\,\,\,\,\,\,\,\,\,\,\,\,\,\,\,\,\,\,\,\,\,\,\,\,\,\,\,\,\,\,\,\,\,\,\,\,\,\,\,\,\,\,\,\,\,\,\,\,\,\,\,\,\,\,\,\,\,\,\,\,\,\,\,\,  n<0,
\end{cases}
\end{align}
where ${\bf{H}} \triangleq {\bf{H}}_{11}$ and ${\bf v}^{(n)}\triangleq{\bf{H}}{\bf s}^{(n)}$.

We allow the integer part of the \ac{to}, $d$, take  values from a set $\mathcal{D}= \{-n_{\rm{s}}+1,\cdots,-1,0,1,\cdots,n_{\rm{s}}-1\}$. Note that the positive values of the delay, $d$, corresponds to situations when the receiver starts early to receive samples. That is, for $d > 0$, the receiver receives  $d$ noise samples from the environment, and then receives the transmitted \ac{ofdm}  samples starting from the ($d +1$)-th sample. Similarly, when $d \leq 0$, the receiver misses the first  $\lvert d \rvert$ samples from the transmitted \ac{ofdm} samples. Allowing $d$ to take both negative and positive values enables the final estimator to be employed for both frame and symbol synchronization. 

The problem of \ac{to} estimation can be formulated as a multiple hypothesis testing problem. Let ${\rm{H}}_d$ denote the hypothesis corresponding to TO $d \in \mathcal{D}=\{-n_{\rm{s}}+1, \cdots,  n_{\rm{s}}-1\}$.
For $d = 0$, the ${\bf y}^{(n)}$ in Eq. \eqref{eq : siso matrix form conv 2} is statistically periodic with period $n_{\rm{s}}$ since the signal samples and the noise samples are all statistically periodic. Statistically periodic here means that the \ac{pdf} of the individual received samples repeats itself after certain number of samples. This means that the moments of the samples are also periodic with period $n_{\rm{s}}$. Hence, when $d = 0$, we only need to derive the moments of the samples for the first $n_{\rm{s}}$ samples and repeat them in the same order to obtain the moments of subsequent samples. 

The periodic pattern in the moments of ${\bf y}^{(n)}$ given $H_0$   makes the \ac{mom}  an efficient way for \ac{to} estimation. 
 The main idea behind the \ac{mom} is to derive a theoretical expression for one a moment of the receive signal given different hypothesises and compare it with the sample mean estimate of the moment. 

The \ac{fom} of the received samples in \eqref{eq : siso matrix form conv 2} is  zero, i.e. $\mathbb{E} \big\{ {\bf y}^{(n)} \big\}=0$, since the expected value of the channel taps, the \ac{ofdm} signal samples, and the noise samples are all zero. Hence, the \ac{fom} of the received samples cannot be used for TO estimation. Moreover, the derivation of a theoretical expression for the  higher order moments of the received signal in \eqref{eq : siso matrix form conv 2} is challenging because of the coupling between the channel and signal.
Therefore, we aim at obtaining the \ac{som}  of the received samples. Theorem \ref{theo : power} derives the \ac{som}
of the received samples for a single \ac{ofdm} block, i.e. samples with indices from 0 to $n_{\rm{s}}-1$ given $H_0$.
For the simplicity of

\begin{theorem} \label{theo : power}
The \ac{som} of the received \ac{ofdm} samples given hypothesis $H_0$  are given by
\begin{equation} \label{eq : power}
M_0[k] = M_0[k+nn_{\rm s}] \triangleq \mathbb{E}
  \big\{ \vert y^{(n)}[k] \vert^2 |{\rm{H}}_0 \big\} =  \sigma^2_{\rm s} \sum_{l = a_k}^{b_k} \sigma^2_{h_l} 
+ \sigma^2_{\rm n} 
\end{equation}
where $n \in \mathbb{Z}_{\geq 0}$, $k=0, 1,\ldots n_{\rm s}-1$, 
and 
 \begin{equation} \label{eq: a b range}
(a_k,b_k)  \hspace{-0.2em}= \hspace{-0.2em}
     \begin{cases}
      (0,k) &~ 0 \le k \le n_{\rm{h}}-2\\
       (0,n_{\rm{h}}-1) &~ n_{\rm{h}}-1 \le k \le n_{\rm{x}}-1\\
       (k-n_{\rm{x}}+1,n_{\rm{h}}-1) &~ n_{\rm{x}} \le k \le n_{\rm{x}}+n_{\rm{h}}-2\\
       ({\rm NaN}, {\rm NaN})  &~ n_{\rm{x}}+n_{\rm{h}}-1 \le k \le n_{\rm{s}}-1.
     \end{cases}
\end{equation}
with $n_{\rm s} = n_{\rm x}+n_{\rm z} -1 $
and $n_{\rm h} \leq n_{\rm z}$. 
\end{theorem}
\begin{proof}
Considering the fact that signal and noise are independent and by using \eqref{eq : siso matrix form conv 2} for $n \ge 0$, we have 
\begin{equation} \label{eq : proof power}
\mathbb{E}
\big\{ \vert {\bf y}^{(n)}[k] \vert^2 |{\rm{H}}_0 \big\} =   \mathbb{E}
\big\{ \vert {\bf v}^{(n)}[k] \vert^2  \big\} + \mathbb{E}
\big\{ \vert
{\bf w}^{(n)} [k] \vert^2  \big\}.
\end{equation}
The expansion of ${\bf v}^{(n)} = {\bf{H}}{\bf s}^{(n)}$ is given in Equation \eqref{eq : expa conv 2}. Taking the expected value of Equation \eqref{eq : expa conv 2} and substituting the result in   \eqref{eq : proof power} yields Equations \eqref{eq : power} and \eqref{eq: a b range}.

\begin{figure*}[t] 
\label{eq:6767}

 \begin{equation} \label{eq : expa conv 2}
v^{(n)}_{{\rm{I}}}[k] \hspace{-0.2em}= \hspace{-0.2em}
     \begin{cases}
      \sum^{m}_{u=0} h_{\rm{I}}[k,u] s^{(n)}_{{\rm{I}}}[k-u] - h_{\rm{Q}}[k,u] s^{(n)}_{{\rm{Q}}}[k-u]  ~~~~~~~~~~~~~~~~~~~~~ 0 \le k < n_{\rm{h}}-2 \\
       \sum^{n_{\rm{h}}-1}_{u=0} h_{\rm{I}}[k,u] s^{(n)}_{{\rm{I}}}[k-u] - h_{\rm{Q}}[k,u] s^{(n)}_{{\rm{Q}}}[k-u]  ~~~~~~~~~~~~~~~~~~~~ n_{\rm{h}}-1 \le k \le n_{\rm{x}}-1, \\
       \sum^{n_{\rm{h}}-1}_{u=m-n_{\rm{x}}+1} h_{\rm{I}}[k,u] s^{(n)}_{{\rm{I}}}[k-u] - h_{\rm{Q}}[k,u] s^{(n)}_{{\rm{Q}}}[k-u]  ~~~~~~~~~~~~~ n_{\rm{x}} \le k \le n_{\rm{x}}+n_{\rm{h}}-2,\\
       0 ~~~~~~~~~~~~~~~~~~~~~~~~~~~~~~~~~~~~~~~~~~~~~~~~~~~~~~~~~~~~~~~~~~~~~~~~~~~~ n_{\rm{x}}+n_{\rm{h}}-1 \le k \le n_{\rm{s}}-1,
     \end{cases}
\end{equation}
\\
\noindent\rule{\textwidth}{1pt}
\end{figure*}

\end{proof}

Note that the right hand side of Eq. \eqref{eq : power} is independent of $n$, and only depends on $k$. Hence,  $M_0[k], \forall k \in \mathbb{N}_{\geq 0}$ is periodic with period $n_{\rm{s}}$.

Now, we have the theoretical expression for the $n_{\rm s}=n_{\rm x}+ n_{\rm h}-1$ \ac{som}s of an \ac{ofdm} block. In order to estimate the \ac{to}, $d$, we need to obtain the corresponding sample mean estimate of the \ac{som}s and compare them 
 with the corresponding theoretical values in  \eqref{theo : power}. 
 In the following, we first obtain the sample mean estimate of the \ac{som}s of an \ac{ofdm} block and then formulate the \ac{to} estimation as \ac{ls} minimization problem.  

\section{TO Estimation for \ac{siso}-\ac{ofdm}}
In this section, we propose \ac{som} and weighted \ac{som} \ac{to} estimators for \ac{zp}-\ac{ofdm}. 

\subsection{\ac{som} Estimation} \label{subsec : finding moment}
Let us consider an observation window of length $L$ and  define the observation vector ${\bf y}_{\rm rec}$ as 
\begin{align}
{\bf y}_{\rm rec} \triangleq \big{[} y_{\rm rec}[0] \  y_{\rm rec}[1] \ \ldots \ y_{\rm rec}[L-1]  \ \big{]}^{\rm T}. 
\end{align}
In order to estimate the TO given ${\bf y}_{\rm rec}$ by using the \ac{mom}, we need to define the sample mean estimate of two conditional \ac{som}s in this subsection. 

Because  $M_0[k]$ is statistically periodic, 
the sample mean estimate of $\mathbb{E}
  \big\{ \vert y^{(n)}[k+d] \vert^2 {\rm{H}}_d \big\}=M_0[k]$ for 
$k = 0,1,\ldots,n_{\rm s}-1$  
and $d > 0$ is given by 
\begin{align} \label{eq : positive practical power}
\nonumber \hat{M}^{+}_{ d, L }[k] 
\triangleq& 
~
\frac{1}{\lfloor (L - k -  d  + 1)/n_{\rm{s}} \rfloor} \times 
\\ \nonumber & 
~~~~\sum_{r = 0}^{ \lfloor (L - k - d  + 1)/n_{\rm{s}} \rfloor - 1} \vert y_{\text{rec}}[k +  d  + r n_{\rm{s}}] \vert^2, \\  
&\forall k \in \{0, 1, \cdots , n_{\rm{s}}-1 \}, 
\end{align}
where $+$ subscript denotes $d >0$.
Given hypothesis $H_d$, $d >  0$, the noise variance can be estimated as 
\begin{equation} \label{eq : noise practical power}
\hat{M}^{\text{noise}}_{d}  =     \frac{1}{ d } \sum_{r = 0}^{  d  - 1 } \vert y_{\text{rec}}[r] \vert^2.
\end{equation}

Similarly, the sample mean estimate of $\mathbb{E}
  \big\{ \vert y^{(n)}[k] \vert^2 {\rm{H}}_d \big\}=M_0[k+|d|]$ for $k = 0,1,\ldots,n_{\rm s}-1$   and $d \leq 0$ is given by 
\begin{align} \label{eq : negative practical power}
\nonumber \hat{M}^{-}_{d, L}[k] =& 
~
\\=& 
\frac{1}{\lfloor (L - k + 1)/n_{\rm{s}} \rfloor } \sum_{r = 0}^{ \lfloor (L - k + 1)/n_{\rm{s}} \rfloor - 1} \vert y_{\text{rec}}[k + r n_{\rm{s}}] \vert^2, \\  \nonumber
&\forall k \in \{0, 1, \cdots , n_{\rm{s}}-1 \}, 
\end{align}
where subscript $-$ denotes $d \leq 0$.

In the asymptotic case that $L \rightarrow +\infty$, we have 
\begin{align} \label{eq : inf mapping equality > 0}
 \nonumber &\lim_{L \to \infty} ~\hat{M}^{+}_{ d, L }[k] \mapsto M_0[k],
\\  
&~~~~~~~~ \forall k \in \{0,  1, \cdots , n_{\rm{s}} - 1 \}
\end{align}
and 
\begin{align} \label{eq : inf mapping equality < 0}
\lim_{L \to \infty} ~ \hat{M}^{-}_{ d, L }[k] &\mapsto M_0[k + \lvert d \rvert ], 
\\  \nonumber 
&\forall k \in \{0, 1, \cdots , n_{\rm{s}}-1 \}.
\end{align}

\begin{figure}
\centering
\includegraphics[height=2.835in]{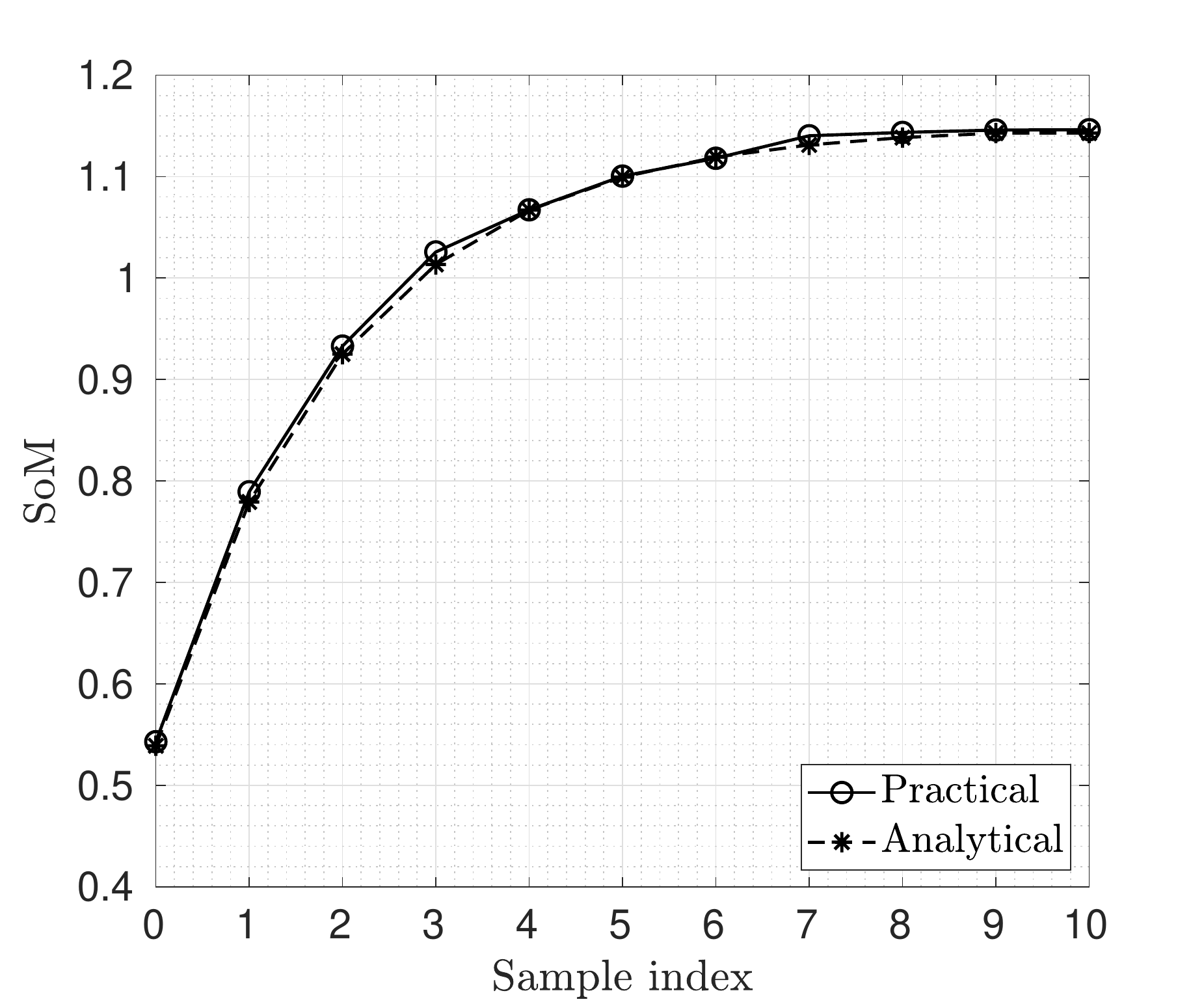}
  \caption{The practical and analytical \ac{som} of the starting samples of an \ac{ofdm} symbol.} \label{fig: som_start}
  \vspace{-1em}
\end{figure}

\begin{figure}
\centering
\includegraphics[height=2.835in]{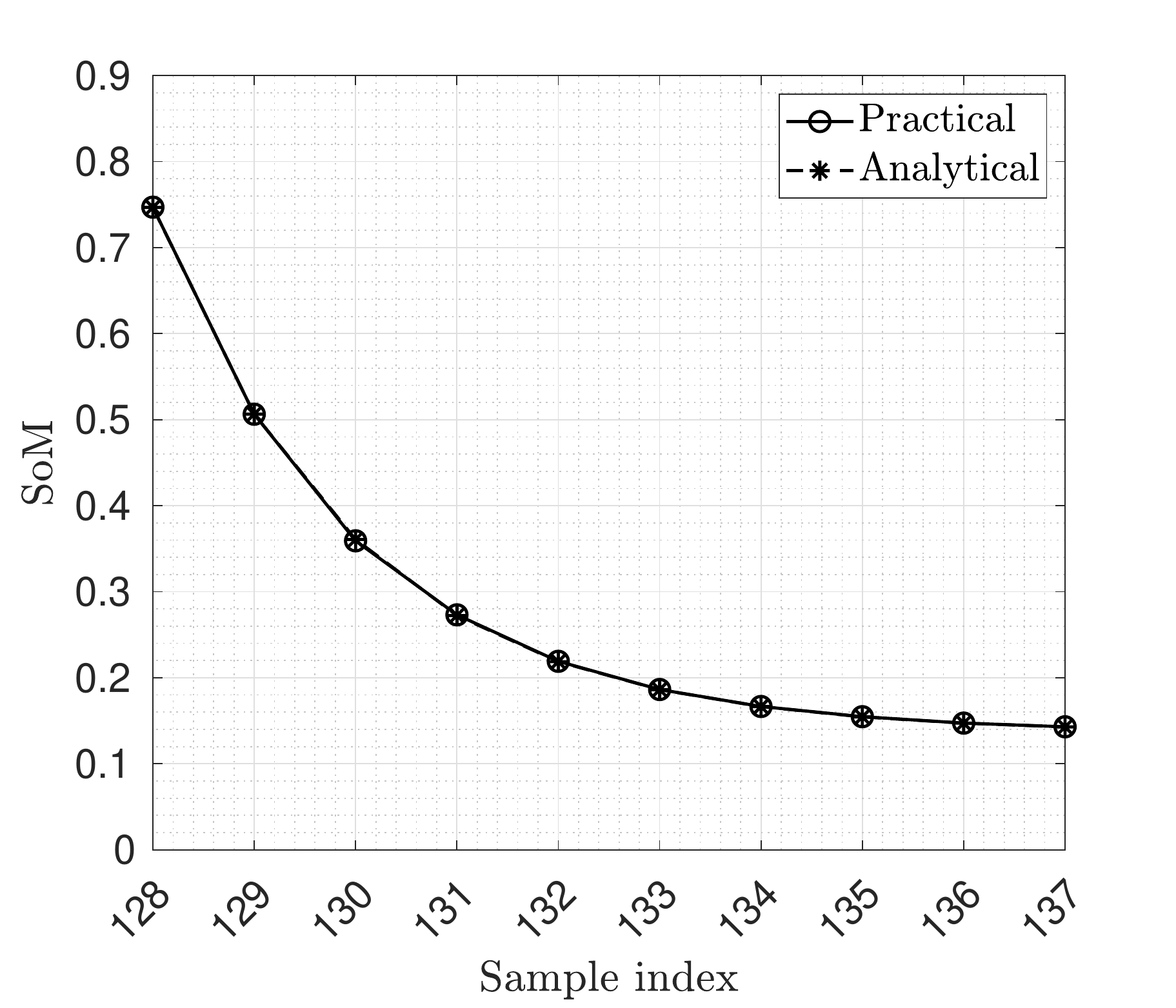}
  \caption{The practical and analytical \ac{som} of the trailing samples of an \ac{ofdm} symbol.} \label{fig: som_end}
  \vspace{-1em}
\end{figure}

Figs \ref{fig: som_start} and \ref{fig: som_end} show the asymptotic convergence of the \ac{som}s versus the index of the samples in an \ac{ofdm} symbol with $10^5$ Monte Carlo realizations. 


\subsection{\ac{to} Estimation} 

\label{subsection : mapping}

In the asymptotic case, $\hat{d}$ is $d \in \mathcal{D}$ that satisfies all the 

to estimate \ac{to} $d$, we need to iterate over different values of $d \in \mathcal{D}$, and find the one that 
satisfies all the equations 
makes the set of equalities in either \eqref{eq : inf mapping equality > 0} \big(and \eqref{eq : noise practical power}\big) or \eqref{eq : inf mapping equality < 0} hold. However, since the number of the received samples $L$ at the receiver is finite, the equalities in Equations \eqref{eq : inf mapping equality > 0} \footnote{and \eqref{eq : noise practical power}, since $d$ is also finite.} and \eqref{eq : inf mapping equality < 0} turn into semi-equalities meaning that the values would be very close but not equal for an optimal $d$. Thus, we need to compromise here and use the absolute value of differences as follows 

\begin{align} \label{estimator : unweighted}
\min_{d \in \mathcal{D}}    \sum_{k = 0}^{n_{\rm{s}} - 1} &
\Big\lvert \hat{M}^{-}_{ d, L }[k] - M_0[k + \lvert d \rvert ] \Big\rvert ~u[-d] 
\\ \nonumber 
&~ +   \Big\lvert \hat{M}^{+}_{ d, L }[k] - M_0[k] \Big\rvert  u[d - 1] + \Big\lvert \hat{M}^{\text{noise}}_{ d } - \sigma^2_n \Big\rvert u[d - 1]
\end{align}

\noindent where $u[d]$ is discrete unit step function, defined as 
\begin{equation}
u[k] = \begin{cases} 1, & k \geq 0 \\ 0, & k < 0. \end{cases}    
\end{equation}

From Eq. \eqref{estimator : unweighted}, it is clear that the accuracy of the estimator heavily depends on how large the number of the received samples $L$ is. This is a limiting factor in resource-limited devices where a large number of \ac{ofdm} symbols cannot be loaded into memory. Hence, affecting the accuracy of the estimation. In order to address this issue and achieve higher accuracy, we will improve the estimator in  Eq. \eqref{estimator : unweighted} in the next subsection.


\subsection{Weighted \ac{som} estimator for \ac{siso}-\ac{ofdm}}

In this subsection, we propose and develop the  \ac{wsom} estimator that offers a higher probability of lock-in. The main idea of \ac{wsom} estimator is to give each sample's moment a weight which describes the  confidence on the closeness of the analytical and practical means.


The variance of a random variable  approximately shows  how close on average is the realization value of a random value to its mean value. In our case, the mean value is the second moments of the samples we derived in Theorem  \ref{theo : power}, and the variance of those second moments shows how close on average the second moments of the \textit{received} samples, i.e. the left hand sides in Eq. \eqref{eq : inf mapping equality > 0} and \eqref{eq : inf mapping equality < 0} would be to their mean value, i.e. the right hand sides in Eq. \eqref{eq : inf mapping equality > 0} and \eqref{eq : inf mapping equality < 0}. That is, the smaller the variance, the closer the second moments of the received samples, i.e. the left hand sides in Eq. \eqref{eq : inf mapping equality > 0} and \eqref{eq : inf mapping equality < 0}, would be to their mean value, i.e. the right hand sides in Eq. \eqref{eq : inf mapping equality > 0} and \eqref{eq : inf mapping equality < 0}. Hence, the inverse of the variance of $\vert y^{(n)}[k] \vert^2, k = 1,\ldots,n_{\rm s}$  shows our confidence in how close the left and right hand sides of Equations  \eqref{eq : inf mapping equality > 0} and \eqref{eq : inf mapping equality < 0} are. Next Theorem provides the variance of the \ac{som} of the received samples

\begin{theorem} \label{theo : weights}
The variance of the \ac{som}s of different received \ac{ofdm} samples are given as follows
\begin{equation} \label{eq : weights}
\begin{split} 
\sigma^2_{\vert y^{(n)}[k] \vert^2 |{\rm{H}}_0 } = \frac{13}{4} \sigma^4_{s} \sum_{l = a_k}^{b_k} \sigma^4_{h_l} + \frac{5}{2} \sigma^4_{s} \sum_{r = a_k}^{r = b_k} \sum_{l = a_k, l \neq r}^{l = b_k} \sigma^2_{h_l} \sigma^2_{h_r}  
\\ 
+ 2~ \sigma^2_{n} \sigma^2_{s} \sum_{l = a_k}^{b_k} \sigma^2_{h_l} + \sigma^4_{n}  - \frac{1}{4} \sigma^2_{s} \big(\sum_{l = a_k}^{b_k} \sigma^2_{h_l} \big)^2 
\end{split} 
\end{equation}
where $n \in \mathbb{Z}_{\geq 0}$,  
 \begin{equation} \label{eq: expa conv 21}
(a_k,b_k)  \hspace{-0.2em}= \hspace{-0.2em}
     \begin{cases}
      (0,k) &~ 0 \le k \le n_{\rm{h}}-2\\
       (0,n_{\rm{h}}-1) &~ n_{\rm{h}}-1 \le k \le n_{\rm{x}}-1\\
       (k-n_{\rm{x}}+1,n_{\rm{h}}-1) &~ n_{\rm{x}} \le k \le n_{\rm{x}}+n_{\rm{h}}-2.
     \end{cases}
\end{equation}
\noindent and it is equal to $2 \sigma^2_n$ for $n_{\rm{x}}+n_{\rm{h}}-1 \le k \le n_{\rm{s}}-1$.
\end{theorem}

\begin{proof}
See Appendix \ref{proof : weights}.
\end{proof}

\begin{figure}
\centering
\includegraphics[height=2.835in]{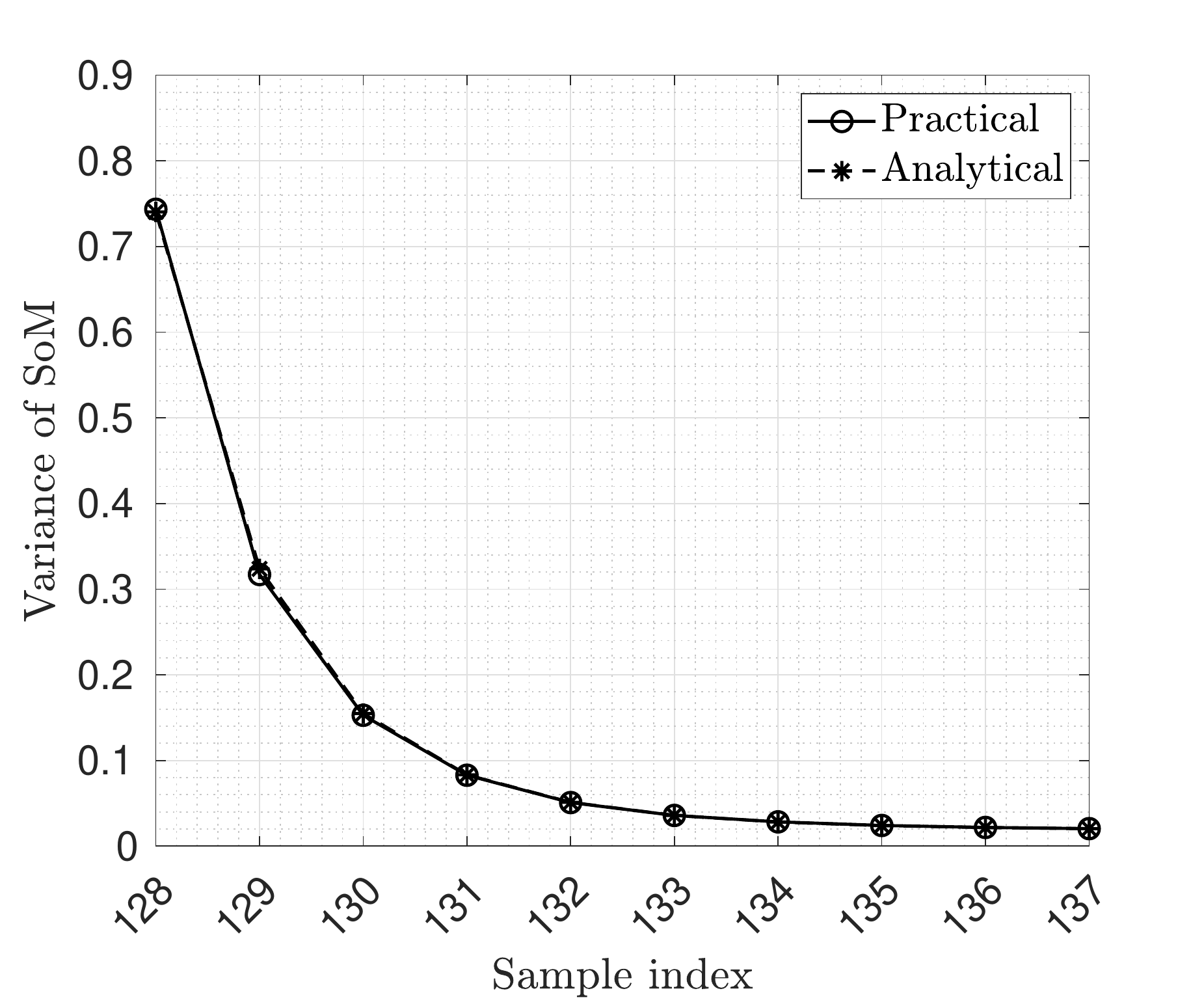}
  \caption{The practical and analytical variance of \ac{som} of the trailing samples for an \ac{ofdm} symbol.} \label{fig: variance_som_end}
  \vspace{-1em}
\end{figure}

Figures \ref{fig: variance_som_start} and \ref{fig: variance_som_end} show the asymptotic convergence of the variance of the \ac{som}s of the starting and trailing samples for an \ac{ofdm} symbol with $10^5$ Monte Carlo realizations.


\begin{figure}
\centering
\includegraphics[height=2.835in]{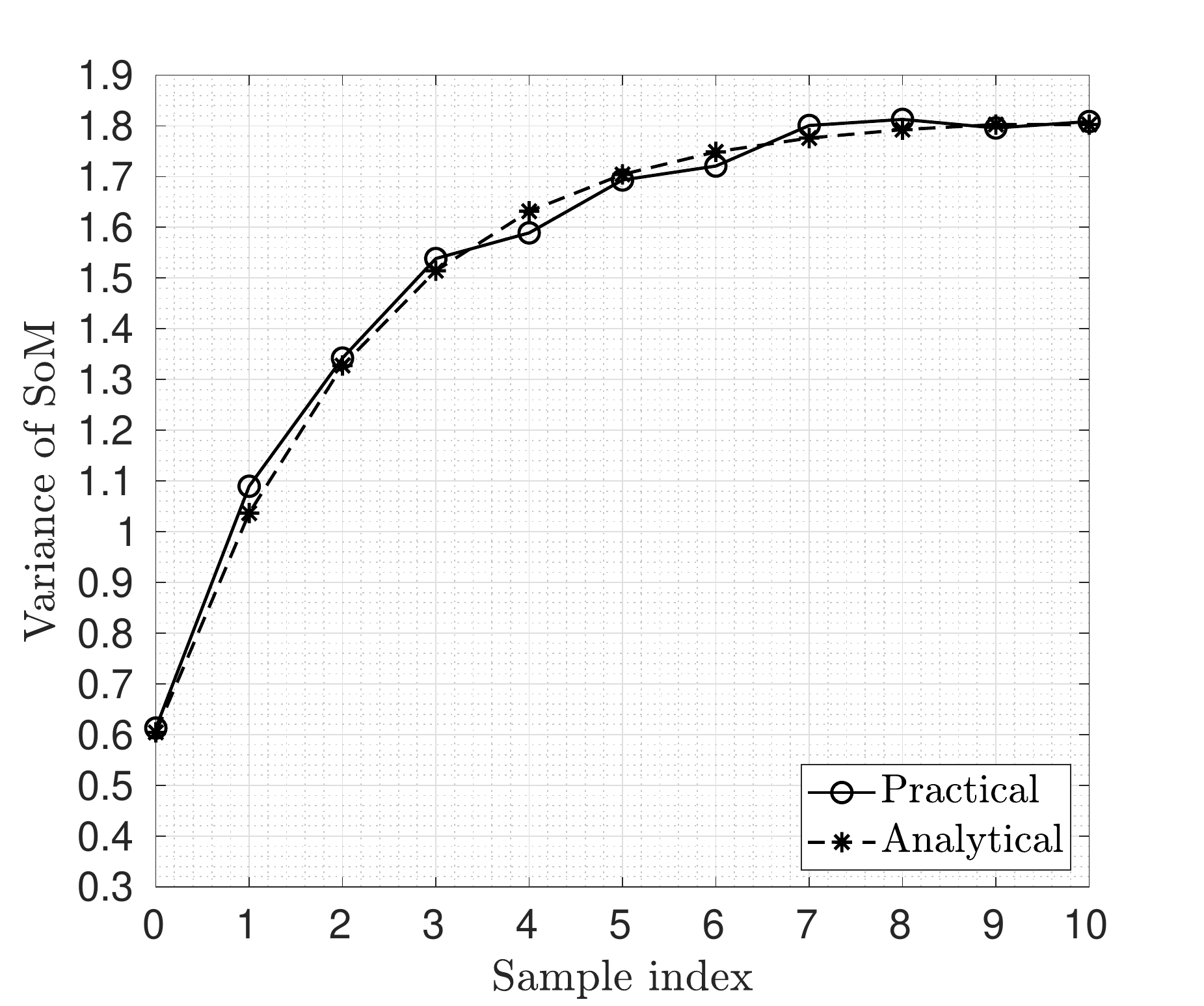}
  \caption{The practical and analytical variance of \ac{som} of the starting samples for an \ac{ofdm} symbol.} \label{fig: variance_som_start}
  \vspace{-1em}
\end{figure}

Now, we  use the inverse of the variance of the \ac{som}s, given in Theorem \ref{theo : weights}, as weights in Eq. \eqref{estimator : unweighted} to improve the accuracy of the estimator. Let us define 

\begin{align} \label{eq : weights vector def}
\nonumber  F[k + n n_{\rm{s}}]
\triangleq& 
~\sigma^2_{\vert y^{(n)}[k] \vert^2 |{\rm{H}}_0 }
\\  
F^{\text{noise}} \triangleq 2 \sigma^2_n,
\\  \nonumber 
&\forall k \in \{0, 1, \cdots , n_{\rm{s}}-1 \},
\\ \nonumber 
&\forall n \in \mathbb{Z}_{\geq 0}.
\end{align}

Clearly, each $F[i]$ maps to $M_0[i]$ defined in Theorem \ref{theo : power}, hence, one can write the weighted version of the estimator in Eq. \eqref{estimator : unweighted} as 
\begin{align} \label{estimator : weighted}
\min_{d \in \mathcal{D}}    \sum_{k = 0}^{n_{\rm{s}} - 1} &
\frac{1}{F[k + \lvert d \rvert ]} \Big\lvert \hat{M}^{-}_{ d, L }[k] - M_0[k + \lvert d \rvert ] \Big\rvert ~u[-d] 
\\ \nonumber 
&~ +   \frac{1}{F[k]} \Big\lvert \hat{M}^{+}_{ d, L }[k] - M_0[k] \Big\rvert u[d - 1]
\\ \nonumber 
&
+ \frac{1}{F^{\text{noise}}} \Big\lvert \hat{M}^{\text{noise}}_{ d } - \sigma^2_n \Big\rvert u[d - 1]
\end{align}

This completes our derivation of the \ac{wsom} estimator. Next, we generalize the derived estimators in \eqref{estimator : unweighted} and \eqref{estimator : weighted} to \ac{mimo} systems.

\subsection{Extension to \ac{mimo} systems}

In this subsection, we assume both the transmitter and the receiver have multiple antennas. For the simplicity of derivations, we keep the total power transmitted through multiple transmit antennas equal to that of single antenna transmitter in \ac{siso} systems. We also assume the delay at different receive antennas are all equal to $d$.

In order to extend the proposed estimators to \ac{mimo} systems, we need to find the accumulative \ac{som} of the received samples at different receive antennas. Let us assume the transmit antennas at the transmitter are used for spatial multiplexing. That is, different signals (information) are sent through different antennas at the transmitter. This means that the \ac{som} of the received samples at the receiver are simply the sum of the \ac{som}s of the received samples coming from each transmit antenna. Using Equations \eqref{Sys Model: matrix form conv 2} and \eqref{matrix H}, we can write
\begin{align} 
\mathbb{E}\big\{ \vert {\bf y}^{(n)}_j[k] \vert^2 |{\rm{H}}_0 \big\} &= (\sum^{m_{\rm t}}_{i = 1} {\bf H}_{ji} {\bf s}^{(n)}_i)[k] + {\bf w}^{(n)}_{j}[k] \\ \nonumber 
&=  \sum^{m_{\rm t}}_{i = 1}  \sum_{l = a_k}^{b_k} \sigma^2_{s_i} \sigma^2_{h_l} + \sigma^2_{n}  \\ \nonumber 
&= \sum^{m_{\rm t}}_{i = 1} \sigma^2_{s_i} \sum_{l = a_k}^{b_k}  \sigma^2_{h_l} + \sigma^2_{n}  \\ \nonumber 
&= \sigma^2_{s} \sum_{l = a_k}^{b_k}  \sigma^2_{h_l}
+ \sigma^2_{n} 
\end{align}
\noindent where the last equality comes from the fact that $\sum^{m_{\rm t}}_{i = 1} \sigma^2_{s_i} = \sigma^2_{s}$, and $a_k$ and $b_k$ are given in \eqref{eq: a b range}. Hence, the theoretical \ac{som}s for each received sample at each antenna $j$ remain the same as the one in Theorem \ref{theo : power}.

Similar to our discussions for \ac{siso} systems, the practical \ac{som}s of the received  samples can be derived as
\begin{align} \label{eq : mimo positive practical power}
\nonumber B^{-}_{ d, L}[k] 
\triangleq& 
~
\frac{1}{m_{\rm{r}} \lfloor (L - k + 1)/n_{\rm{s}} \rfloor} \times 
\\ \nonumber 
&~~~~~~~~~~~~~~~ \sum_{j = 0}^{m_{\rm{r}} - 1} \sum_{r = 0}^{ \lfloor (L - k + 1)/n_{\rm{s}} \rfloor - 1} \vert y^j_{\text{rec}}[k + r n_{\rm{s}}] \vert^2, \\  \nonumber
&\forall k \in \{0, 1, \cdots , n_{\rm{s}}-1 \}. 
\end{align}
\noindent where $y^j_{\text{rec}}$ denotes the vector of the received samples at the $j$-th receive antenna. Similarly, for $d > 0$, one can write 
\begin{align}
B^{\text{noise}}_{ d }   =     \frac{1}{ d } \sum_{j = 0}^{m_{\rm{r}} - 1} \sum_{r = 0}^{ d  - 1 } \vert y^{j}_{\text{rec}}[r] \vert^2.
\end{align}

and 
\begin{align} \label{eq : negative practical power}
\nonumber B^{+}_{ d, L }[k] =& 
\frac{1}{m_{\rm{r}} \lfloor (L - k + 1)/n_{\rm{s}} \rfloor} \times 
\\ \nonumber 
&~~~~~ \sum_{j = 0}^{m_{\rm{r}} - 1} \sum_{r = 0}^{ \lfloor (L - k -   d  + 1)/n_{\rm{s}} \rfloor - 1} \vert y^j_{\text{rec}}[k +   d  + r n_{\rm{s}}] \vert^2, \\  
&\forall k \in \{0, 1, \cdots , n_{\rm{s}}-1 \}.
\end{align}

One can then write the \ac{to} estimator for the \ac{mimo} systems as 
\begin{align} \label{estimator : mimo weighted}
\min_{d \in \mathcal{D}}    \sum_{k = 0}^{n_{\rm{s}} - 1} &
\frac{1}{F[k + \lvert d \rvert ]} \Big\lvert B^{-}_{ d, L }[k] - M_0[k + \lvert d \rvert] \Big\rvert ~u[-d] 
\\ \nonumber 
&~ + \Bigg( \frac{1}{F[k]} \Big\lvert B^{+}_{ d, L }[k] - M_0[k] \Big\rvert 
\\ \nonumber 
&~~~~~~~~~~~~~~~~~~~~~ + \frac{1}{F^{\text{noise}}} \Big\lvert B^{\text{noise}}_{ d  } - \sigma^2_n \Big\rvert \Bigg) u[d - 1]
\end{align}
One can replace $F[i]$s and $F^{\text{noise}}$s with ones to obtain the unweighted estimator for \ac{mimo} systems.  Eq. \eqref{estimator : mimo weighted} as the most general from of the proposed  \ac{to} estimators for \ac{mimo} systems is expanded in Eq. \eqref{estimator : most general form expanded}. 

Deriving an upper bound for the complexity of the \ac{to} estimator proposed in Eq. \eqref{estimator : mimo weighted} is very easy. It should be noted that the weights $F[i]$s and $F^{\text{noise}}$s can be calculated once before any signal reception; hence, the complexity of the \ac{som} estimator in Eq. \eqref{estimator : mimo weighted} is equal to $\mathcal{O}(m_{\rm r} L)$ which is significantly lower than the one in \cite{koosha2020}.


\begin{figure*}[t] 
\label{eq: EXPANDED FINAL MIMO ESTIMATOR}
\begin{align} \label{estimator : most general form expanded}
\min_{d \in \mathcal{D}}    \sum_{k = 0}^{n_{\rm{s}} - 1} & 
\frac{ \Big\lvert \frac{1}{m_{\rm{r}} \lfloor (L - k + 1)/n_{\rm{s}} \rfloor} \sum_{j = 0}^{m_{\rm{r}} - 1} \sum_{r = 0}^{ \lfloor (L - k + 1)/n_{\rm{s}} \rfloor - 1} \vert y^j_{\text{rec}}[k + r n_{\rm{s}}] \vert^2
- \sigma^2_{s} \sum_{l = a_{k+ \lvert d \rvert}}^{b_{k+ \lvert d \rvert}} \sigma^2_{h_l} 
- \sigma^2_{n}  \Big\rvert ~u[-d] }  
{\frac{13}{4} \sigma^4_{s} \sum_{l = a_{k+ \lvert d \rvert}}^{b_{k+ \lvert d \rvert}} \sigma^4_{h_l} + \frac{5}{2} \sigma^4_{s} \sum_{r = a_{k+ \lvert d \rvert}}^{r = b_{k+ \lvert d \rvert}} \sum_{l = a_{k+ \lvert d \rvert}, l \neq r}^{l = b_{k+ \lvert d \rvert}} \sigma^2_{h_l} \sigma^2_{h_r}  
+ 2~ \sigma^2_{n} \sigma^2_{s} \sum_{l = a_{k+ \lvert d \rvert}}^{b_{k+ \lvert d \rvert}} \sigma^2_{h_l} + \sigma^4_{n}  - \frac{1}{4} \sigma^2_{s} \big(\sum_{l = a_{k+ \lvert d \rvert}}^{b_{k+ \lvert d \rvert}} \sigma^2_{h_l} \big)^2 } 
\\ \nonumber 
&~ +  \frac{\Big\lvert \frac{1}{m_{\rm{r}} \lfloor (L - k + 1)/n_{\rm{s}} \rfloor}  \sum_{j = 0}^{m_{\rm{r}} - 1} \sum_{r = 0}^{ \lfloor (L - k -   d  + 1)/n_{\rm{s}} \rfloor - 1} \vert y^j_{\text{rec}}[k +   d  + r n_{\rm{s}}] \vert^2
- \sigma^2_{s} \sum_{l = a_k}^{b_k} \sigma^2_{h_l} 
- \sigma^2_{n}  \Big\rvert  ~u[d - 1] }
{\frac{13}{4} \sigma^4_{s} \sum_{l = a_k}^{b_k} \sigma^4_{h_l} + \frac{5}{2} \sigma^4_{s} \sum_{r = a_k}^{r = b_k} \sum_{l = a_k, l \neq r}^{l = b_k} \sigma^2_{h_l} \sigma^2_{h_r}  
+ 2~ \sigma^2_{n} \sigma^2_{s} \sum_{l = a_k}^{b_k} \sigma^2_{h_l} + \sigma^4_{n}  - \frac{1}{4} \sigma^2_{s} \big(\sum_{l = a_k}^{b_k} \sigma^2_{h_l} \big)^2 } 
\\ \nonumber 
&~~~~~~~~~~~~~~~~~~~~~ + \frac{1}{2 \sigma^2_n} \Big\lvert \frac{1}{ d } \sum_{j = 0}^{m_{\rm{r}} - 1} \sum_{r = 0}^{  d  - 1 } \vert y^{j}_{\text{rec}}[r] \vert^2 - \sigma^2_n \Big\rvert  u[d - 1]
\end{align}
\\
\noindent\rule{\textwidth}{1pt}
\end{figure*}

\section{Simulation results}
\label{sec : simulation results}

In this section, we investigate the performance of the proposed \ac{som} and its weighted version through extensive simulation results under various scenarios.

As shown in Appendix \textcolor{red}{\ref{??????????}}, one can easily show that

\subsection{Simulation Setup}

We consider a \ac{zp}-OFDM system in a doubly-selective (frequency and time) Rayleigh channel. 128-QAM modulation scheme is used for data transmission. If otherwise not specified, the number of data symbols is $n_{\rm{x}}=128$ and the number of zero-padding samples are $n_{\rm{z}}=12$. The number of received \ac{ofdm} symbols, i.e. $N$, used for estimation is 10. The transmit power is set to  {$\sigma_{\rm{x}}^2=1$}.

The sampling frequency at the receiver is set to $f_{\rm sa}=10^9$. A multipath fading channel with maximum delay spread of {$\tau_{\rm{max}}=10$ns}, equivalent to $n_{\rm{h}}=10$ taps, is assumed. An exponential decaying function, i.e., $\sigma^2_{h_l}=\alpha\exp(-\beta l)$, $l=0,1,\dots,n_{\rm{h}}-1$, where  $p_{\rm{h}}=\sum_{l=0}^{n_{\rm{h}}-1}\sigma^2_{h_l}=1$,  $\alpha=1/2.5244$, and $\beta=0.5$, is considered for the delay profile of the fading channel. The maximum Doppler spread of the channel is assumed to be $f_{\rm{D}}=150$ Hz.

An \ac{awgn} is considered and modeled as a zero-mean complex Gaussian random variable with variance $\sigma^2_{\rm{w}}$ derived based on the value of $E_{\rm{b}}/N_0$. The \ac{to} affecting the system is modeled as an integer random variable uniformly distributed in the range $d \in [-30 ,  \ 30]$. The number of Monte Carlo realization is set to $10^4$ for all simulation setups. A strict performance measure, i.e. lock-in probability, is used to measure the effectiveness of the proposed estimators. Lock-in probability here refers the probability that the estimated \ac{to} is equal to the actual \ac{to}. Hence, any non-zero estimation error is considered a missed estimation.

\subsection{Simulation Results}

\begin{figure}
\centering
\includegraphics[height=2.835in]{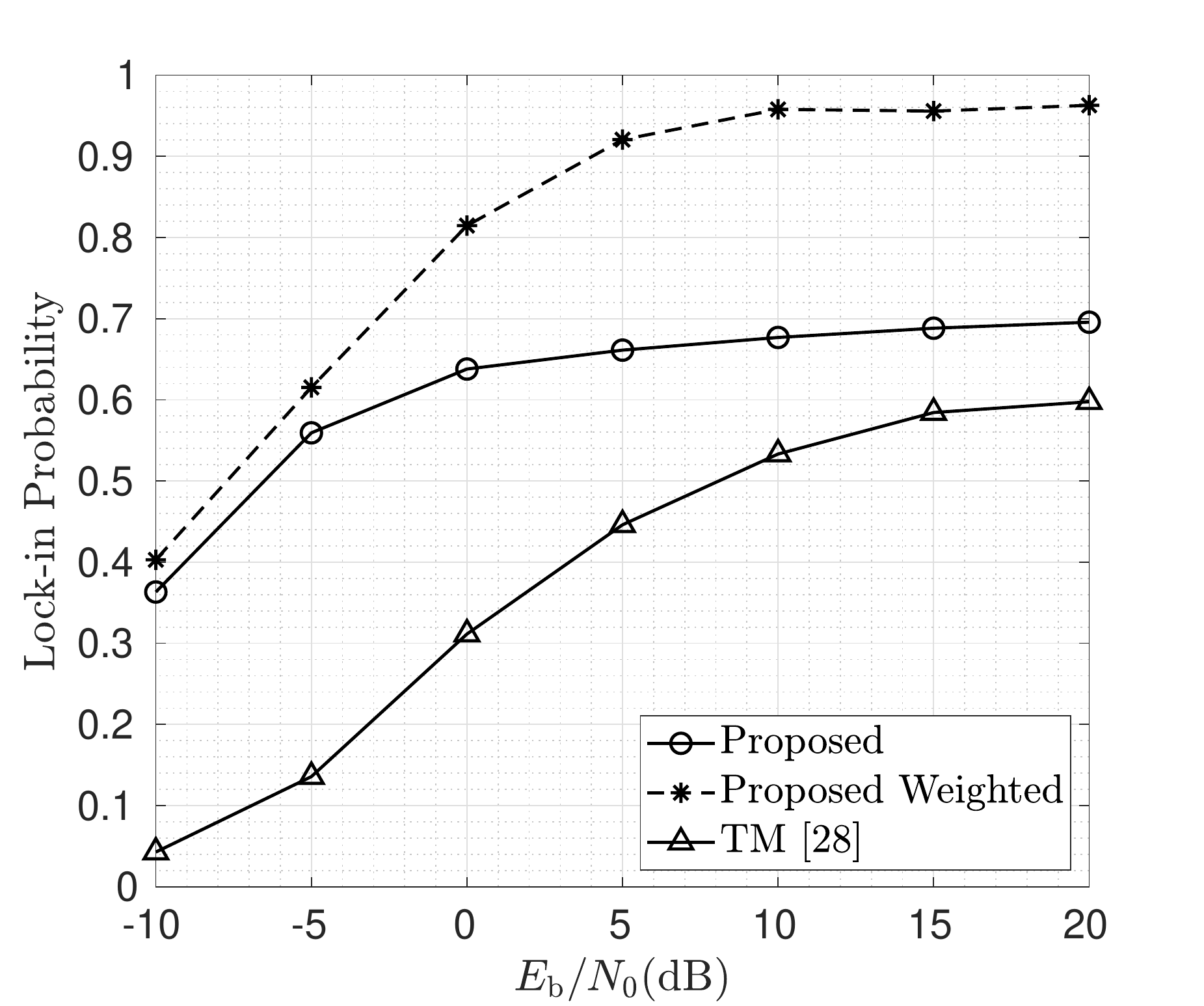}
  \caption{Lock-in probability versus $E_{\rm{b}}/N_0$.}\label{fig: snr}
\end{figure}

We compare the performance of the proposed \ac{som} estimator and its improved weighted version versus the current state-of-the-art estimator proposed by other authors in \cite{LeNir2010}, i.e. transition metric \ac{tm}. The lock-in probability of the proposed estimators versus \ac{tm} for different values of $E_{\rm{b}}/N_0$ are shown in Fig.~\ref{fig: snr}. As can be seen, there is a performance gap between the proposed estimators and \ac{tm} while weighted \ac{som} having the highest probability of lock-in amongst all. This is due to the fact that weighted \ac{som} takes advantage of the statistical information of the channel. The performance of the weighted \ac{som} improves as $E_{\rm{b}}/N_0$ increases because our confidence, modeled as the weights, in the derived theoretical second-order moment, and its closeness to the actual second-order moment of each received sample increases. However, the importance of the proposed estimators manifest itself in lower \ac{snr}s, i.e. 0 to -10 dB, where the lock-in probability is still at 40\% compared to \ac{tm} with negligible lock-in probability. Also, note that unlike \cite{koosha2020}, the proposed estimators can perform for values under 5dB.

\begin{figure}
\centering
\includegraphics[height=2.835in]{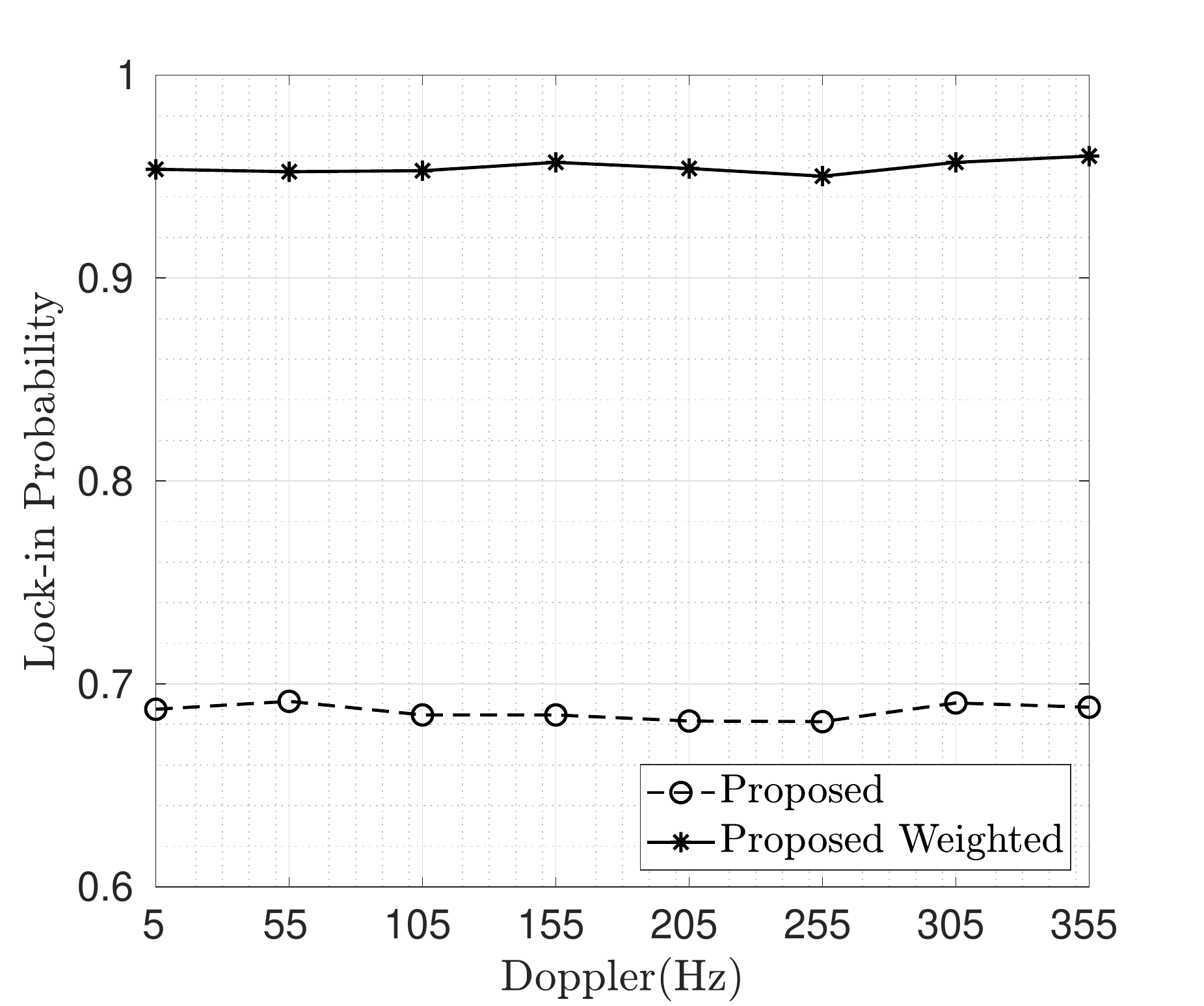}
  \caption{Lock-in probability versus maximum Doppler spread of the fading channel at $15$ dB $E_{\rm{b}}/N_0$.} \label{fig: dop}
  \vspace{-1em}
\end{figure}

The effect of the maximum Doppler spread in Hz (speed of mobility) on the performance of the proposed estimators are shown in Fig. \ref{fig: dop}. As seen, the proposed estimators, unlike \cite{koosha2020}, are relatively independent of the maximum Doppler shift (movement speed). This is mainly due to the fact that unlike \cite{koosha2020} where the joint \ac{pdf} of the received samples is approximated via independency assumption between channel taps, the proposed estimators do not heavily rely on the independency of different channel taps. This makes the proposed estimators an appealing solution for very low to very high speed applications.

\begin{figure}
\centering
\includegraphics[height=2.835in]{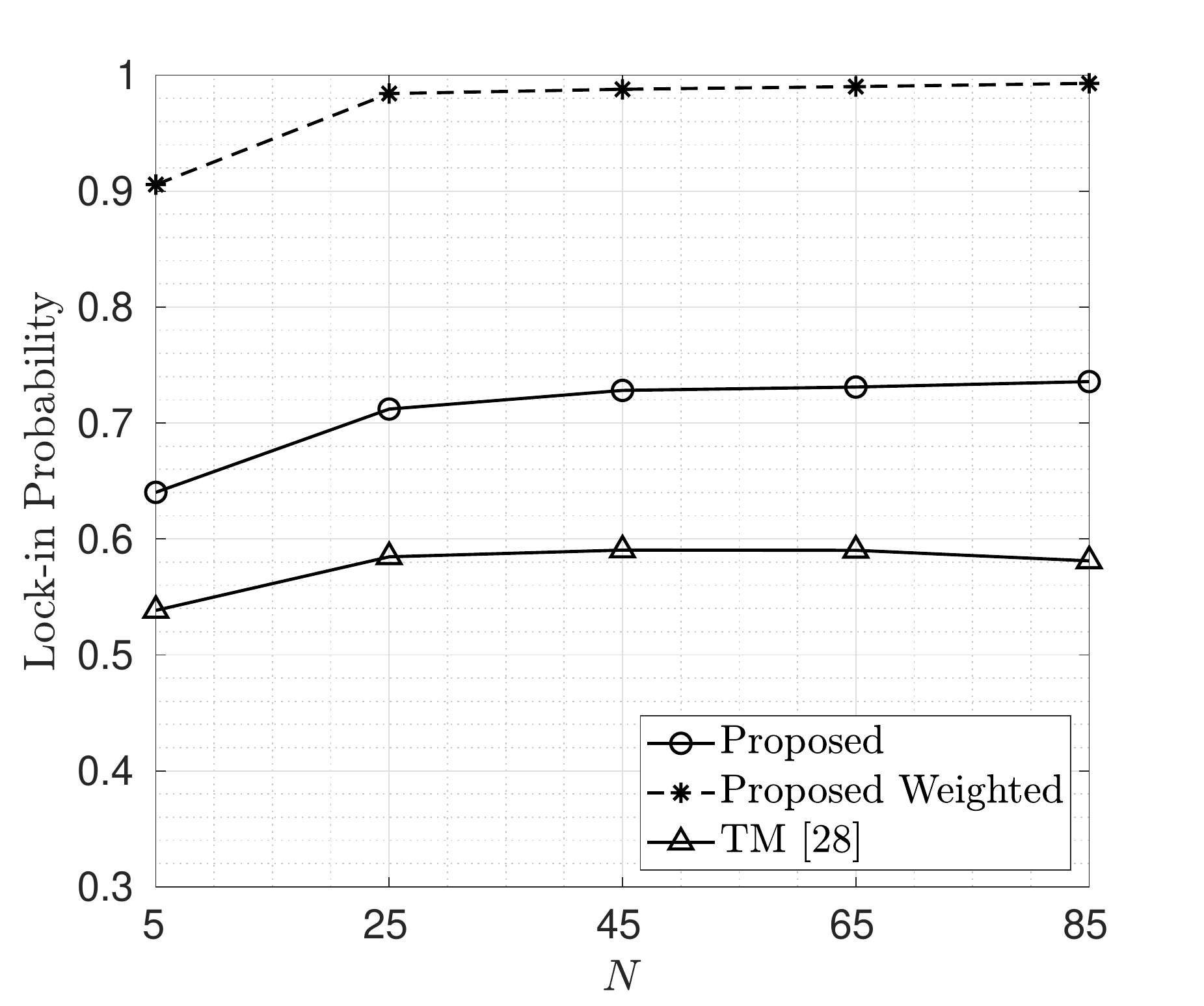}
  \caption{Lock-in probability versus the number of observation vectors $N$ at $15$ dB $E_{\rm{b}}/N_0$.}\label{fig: obs}
\end{figure}

Figure \ref{fig: obs} illustrates the  effect of the number of the observation \ac{ofdm} symbols used for estimation, on the performance of the proposed estimators. As expected, the probability of lock-in increases as the number of observation samples used for estimation increases. The improvements, however, decreases as the number of observation symbols increases. This is because the new information that each new sample adds to the second-order moment and its variance decreases as the number of samples grows.

\begin{figure}
\centering
\includegraphics[height=2.835in]{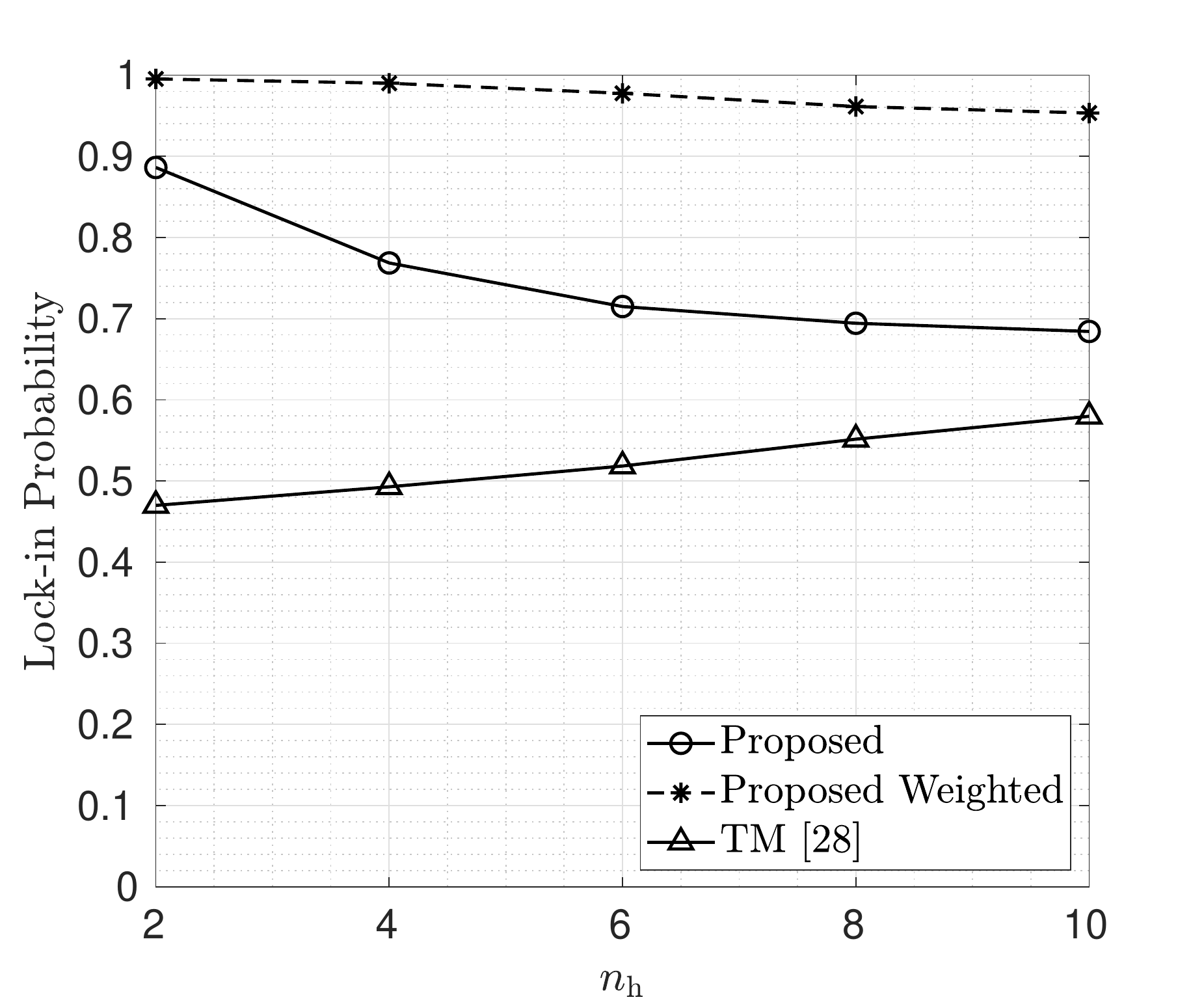}
  \caption{Lock-in probability versus the number of channel taps $n_{\rm{h}}$ at $15$ dB $E_{\rm{b}}/N_0$.}\label{fig: tap}
\end{figure}

The performance of the proposed estimators and \ac{tm} for different values of the number of channel taps $n_{\rm{h}}$, is shown in Fig. \ref{fig: tap}. The lock-in probability of the proposed estimators decreases as the number of channel taps increases. This is because the deviation of the theoretical \ac{som} in \eqref{eq : power} from the actual value increases as the number of channel taps increases. As expected, this decrease in lock-in probability is heavier in unweighted version where the only information is used for estimation is \eqref{eq : power}.

\begin{figure}
\vspace{-1em}
\centering
\includegraphics[height=2.835in]{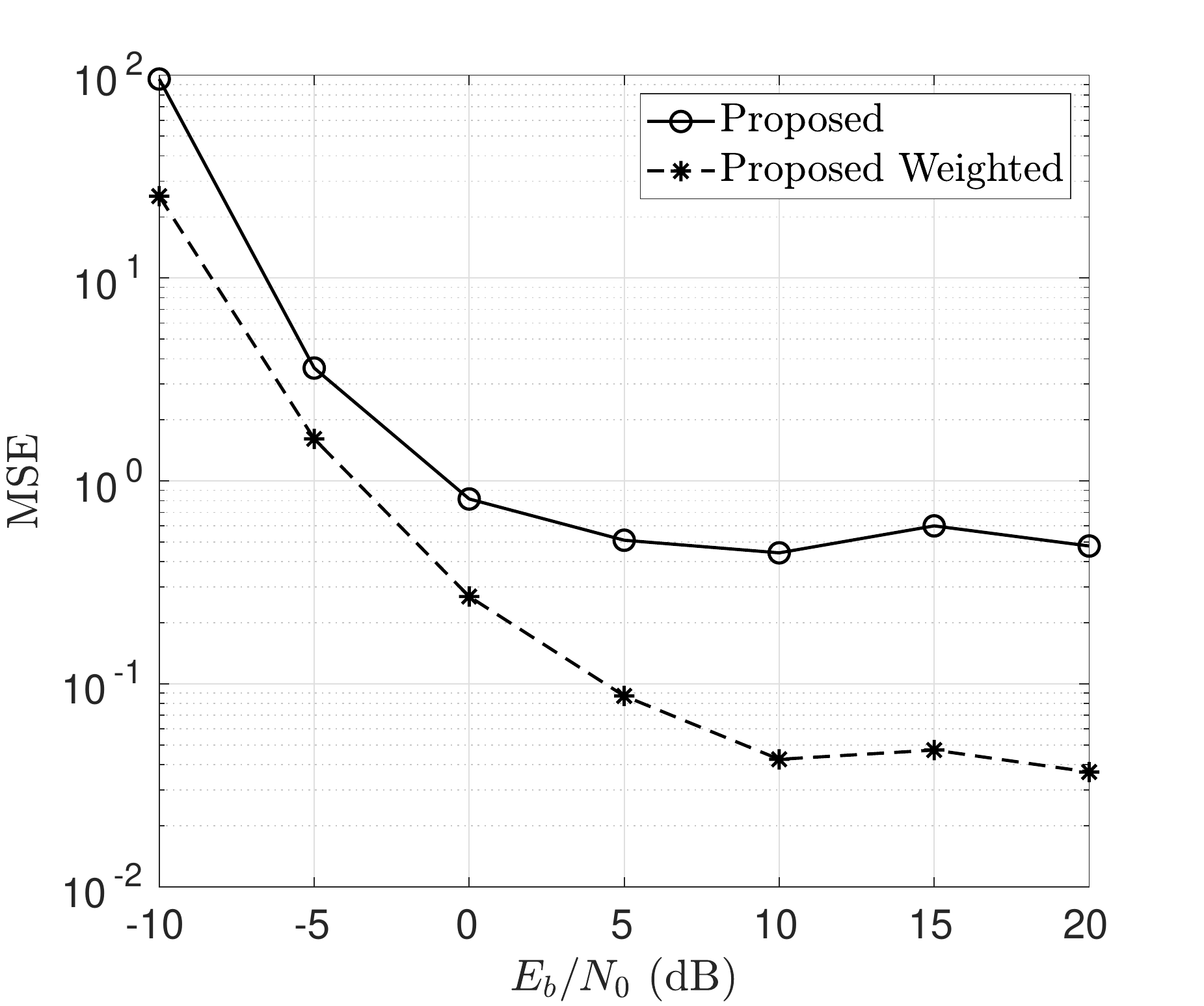}
  \caption{MSE versus $E_{\rm{b}}/N_0$. .}\label{fig: mse}
\end{figure}

\begin{figure}
\centering
\includegraphics[height=2.835in]{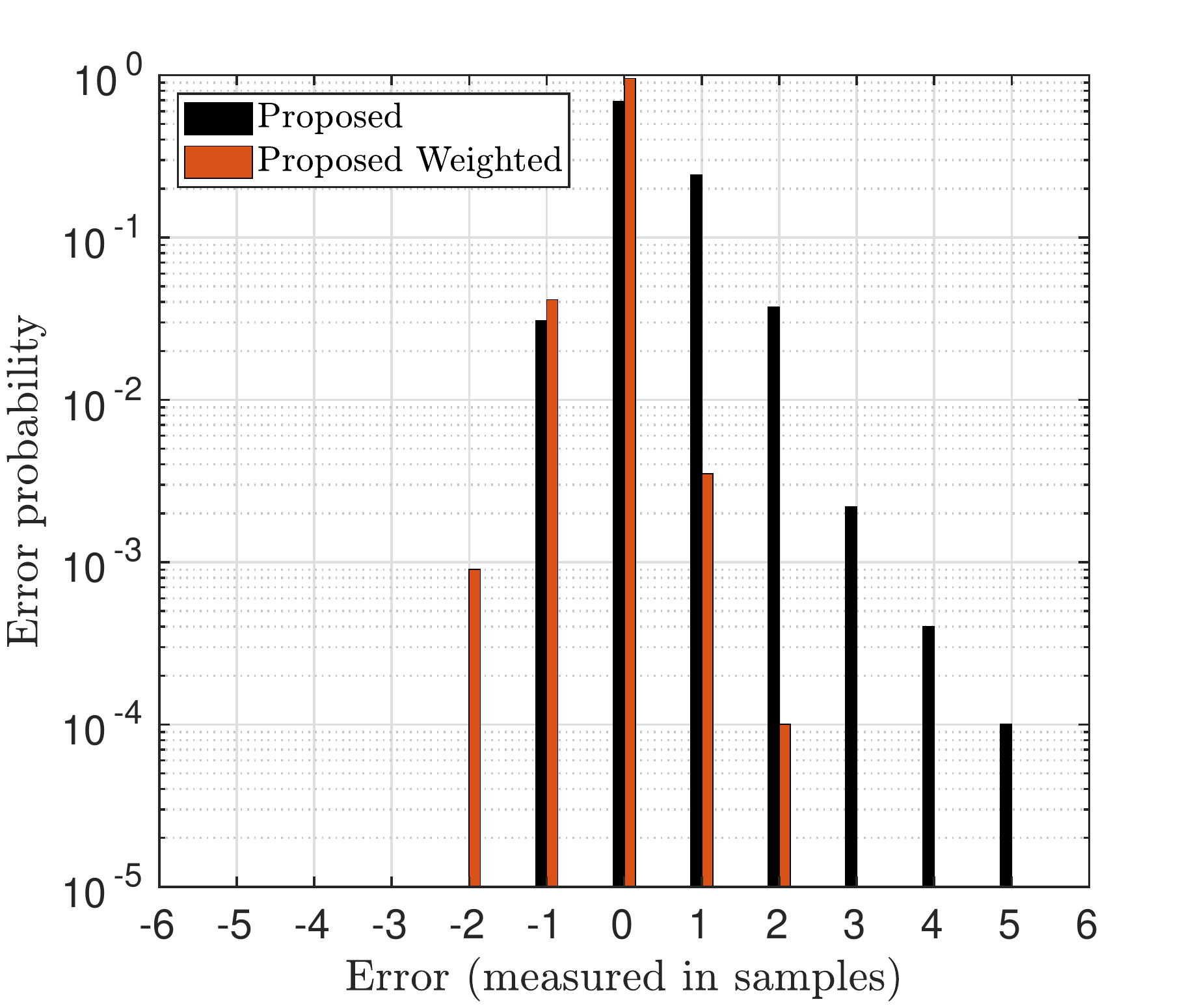}
  \caption{PMF of the synchronization error at 15 dB  $E_{\rm{b}}/N_0$.}\label{fig: hist}
\end{figure}

The empirical probability mass function (PMF) of the synchronization error for the proposed estimators are shown in Fig.~\ref{fig: hist}. As seen, the PMF of the error for weighted \ac{som} is relatively unbiased, while the unweighted \ac{som} is biased towards positive values. Notice, however, that the estimation error for weighted \ac{som} is limited to a range of maximum two neighbouring samples from each side which results in low \ac{mse} as shown in Fig.~\ref{fig: mse}. Fig.~\ref{fig: mse} also shows an average estimation error of less than two for very low \ac{snr}, i.e. -5 dB, for the weighted \ac{som}. Hence, simple and modified versions of the weighted \ac{som} where neighbouring samples of the estimated \ac{to} are also checked can be devised to achieve 100\% lock-in probability.

\begin{figure}
\centering
\includegraphics[height=2.835in]{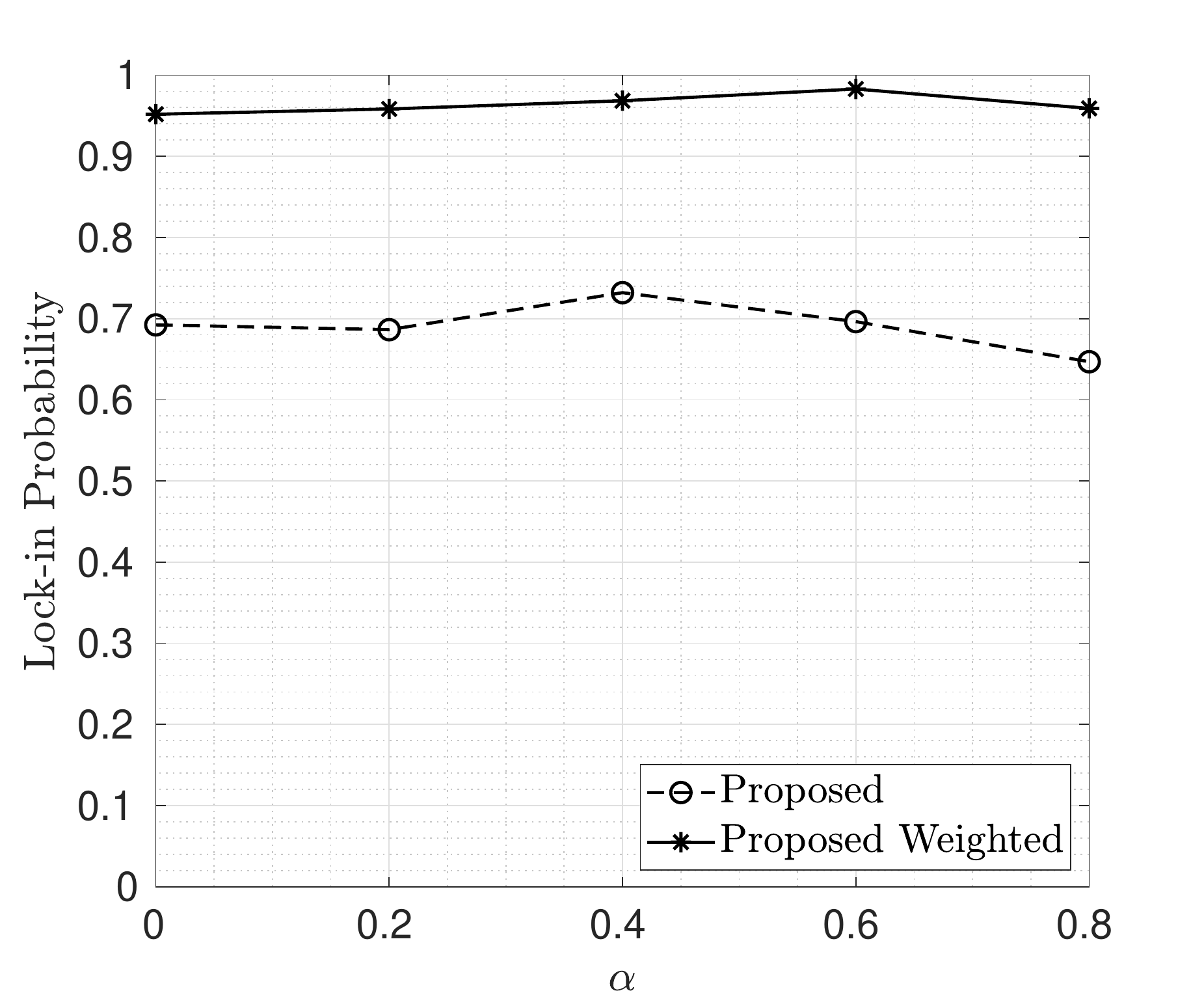}
  \caption{Sensitivity of the proposed synchronization algorithms to PDP estimation error for $N=10$ at 15 dB  $E_{\rm{b}}/N_0$.}\label{fig: sens}
\end{figure}

Fig.~\ref{fig: sens} shows the effect of \ac{pdp} estimation error on the lock-in probability. The estimated \ac{pdp} used for estimation is randomly uniform  set to be either $\sigma^2_{{\rm{h}}_k}-\alpha\sigma^2_{{\rm{h}}_k}$ or $\sigma^2_{{\rm{h}}_k}+\alpha\sigma^2_{{\rm{h}}_k}$ where $\sigma^2_{{\rm{h}}_k}$, $k=0,1,\dots, n_{\rm{h}}-1$, is the true PDP, and $\alpha \in [0,0.8]$ denotes the \ac{pdp} estimation error. As seen, the proposed estimators are relatively robust even for \ac{pdp} estimation errors larger than $50\%$. Note,  however, that the lock-in probability of the weighted \ac{som} remains above $90\%$ even when the the \ac{pdp} estimation error is about $80\%$.

\section{Conclusion}
\label{sec : conclusion}
In this paper, we proposed two low-complexity \ac{nda} \ac{to} estimators based on \ac{som} for \ac{mimo} \ac{zp}-\ac{ofdm} systems in doubly selective channels. Unlike other estimators in the literature, the proposed estimators enjoy very low complexity, are feasible for very low \ac{snr}s, and have very high lock-in probability. Simulation results verify our analysis and demonstrate very high probability of lock-in for the proposed estimators making them suitable for implementation in low-power low complexity devices, i.e. \ac{iot} devices and underwater sensors, unlike the other \ac{nda} \ac{to} estimators for \ac{zp}-\ac{ofdm} systems.


\appendix

\begin{proof} \label{proof : weights}
For notation simplicity and without loss of generality, we  omit the $\hdots|{\rm{H}}_0$ from equations.  For mathematical simplicity, we consider $d = 0$ because non-zero delays are easily derived, mostly via a shift, from zero delay case. We have $\sigma^2_{\vert y_n[k] \vert^2} = \mathbb{E}\{ \vert y_n[k] \vert^4 \} - \mathbb{E}\{ \vert y_n[k] \vert^2 \}^2$. The second term, $\mathbb{E}\{ \vert y_n[k] \vert^2 \}$ has been derived before, and so we need to only focus on $\mathbb{E}\{ \vert y_n[k] \vert^4 \}$. Note that 
\begin{equation} \label{eq: indepency 0}
\begin{split} 
\mathbb{E} \{ \vert y_n[k] \vert^4 \} 
&= 
\mathbb{E} \bigg\{ \big( v^{(n)}[k] + w^{(n)}[k] \big) \big( v^{(n)}[k]^* + w^{(n)}[k]^* \big)  
\\& 
~~~~~~~~ \big( v^{(n)}[k] + w^{(n)}[k] \big) \big( v^{(n)}[k]^* + w^{(n)}[k]^* \big) \bigg\} 
\\&= 
\mathbb{E} \bigg\{  \bigg( | v^{(n)}[k] |^2 + v^{(n)}[k] w^{(n)}[k]^*  +
\\& 
~~~~~~~~~~~~~~~~~~~  v^{(n)}[k]^* w^{(n)}[k] 
+ |w^{(n)}[k]|^2 \bigg)^2 \bigg\}
\\&\stackrel{a}{=} 
\mathbb{E} \bigg\{ | v^{(n)}[k] |^4 + 4 | v^{(n)}[k] |^2 |w^{(n)}[k]|^2 + 
\\& 
~~~~~~~~~~~ |w^{(n)}[k]|^4 + v^{(n)}[k]^2 {w^{(n)}[k]^*}^2 +
\\& 
~~~~~~~~~~~~~~~~~~~ {v^{(n)}[k]^*}^2 w^{(n)}[k]^2 \bigg\}
\\&\stackrel{b}{=} 
\mathbb{E} \Big\{ | v^{(n)}[k] |^4 \Big\} + 4 \mathbb{E} \Big\{ | v^{(n)}[k] |^2 \Big\} \mathbb{E} \Big\{ |w^{(n)}[k]|^2 \Big\}  
\\& 
~~~~~~~~~~~ + \mathbb{E} \Big\{ |w^{(n)}[k]|^4 \Big\}
\end{split}
\end{equation}

\noindent where 
(a) comes from the fact that $v^{(n)}[k]$ and $w^{(n)}[k]$  are independent complex random variables, and $ \mathbb{E}\{ v^{(n)}[k] \} = \mathbb{E}\{ w^{(n)}[k] \} = 0$ and 
(b) is easy to show that $\mathbb{E}\{  v^{(n)}[k]^2 {w^{(n)}[k]^*}^2 + {v^{(n)}[k]^*}^2 w^{(n)}[k]^2 \} = 0$. We have derived $\mathbb{E}\{  |v^{(n)}[k] | ^2 \} $ in previous sections. Also, since $w^{(n)}[k]$ is a complex random variable, we have $\mathbb{E}\{ |w^{(n)}[k]|^2 \} = \sigma^2_w$ and $ \mathbb{E}\{ |w^{(n)}[k]|^4 \} = 2\sigma^4_w$. Hence, the problem further boils down to deriving $\mathbb{E}\{ | v^{(n)}[k] |^4 \}$. After some extensive mathematical manipulation, one can show that 

\begin{equation} \label{eq: indepency 1}
\begin{split} 
&\mathbb{E}\{ | v^{(n)}[k] |^4 \}
\\&= 
\mathbb{E}
\Bigg\{  
\Bigg( 
\bigg( 
\sum^{b}_{u=a} h_{\rm{I}}[k,u] s^{(n)}_{{\rm{I}}}[k-u] - h_{\rm{Q}}[k,u] s^{(n)}_{{\rm{Q}}}[k-u]
\bigg)^2
\\& 
~ +
\bigg( 
\sum^{b}_{u=a} h_{\rm{I}}[k,u] s^{(n)}_{{\rm{Q}}}[k-u] + h_{\rm{Q}}[k,u] s^{(n)}_{{\rm{I}}}[k-u]
\bigg)^2
\Bigg)^2
\Bigg\}
\\&= 
\mathbb{E}
\Bigg\{  
\bigg( 
\sum^{b}_{u=a} h_{\rm{I}}[k,u] s^{(n)}_{{\rm{I}}}[k-u] - h_{\rm{Q}}[k,u] s^{(n)}_{{\rm{Q}}}[k-u]
\bigg)^4
\Bigg\}
\\& 
+
\mathbb{E}
\Bigg\{  
\bigg( 
\sum^{b}_{u=a} h_{\rm{I}}[k,u] s^{(n)}_{{\rm{Q}}}[k-u] + h_{\rm{Q}}[k,u] s^{(n)}_{{\rm{I}}}[k-u]
\bigg)^4
\Bigg\}
\\& 
+ 
2 ~ \mathbb{E}
\Bigg\{  
\bigg( 
\sum^{b}_{u=a} h_{\rm{I}}[k,u] s^{(n)}_{{\rm{I}}}[k-u] - h_{\rm{Q}}[k,u] s^{(n)}_{{\rm{Q}}}[k-u]
\bigg)^2
\\& 
~~~~~~~~~~~
\bigg( 
\sum^{b}_{u=a} h_{\rm{I}}[k,u] s^{(n)}_{{\rm{Q}}}[k-u] + h_{\rm{Q}}[k,u] s^{(n)}_{{\rm{I}}}[k-u]
\bigg)^2
\Bigg\}
\\& \stackrel{(i)}{=} 
2~\mathbb{E}
\Bigg\{  
\bigg( 
\sum^{b}_{u=a} h_{\rm{I}}[k,u] s^{(n)}_{{\rm{I}}}[k-u] - h_{\rm{Q}}[k,u] s^{(n)}_{{\rm{Q}}}[k-u]
\bigg)^4
\Bigg\} 
\\& 
~~~ + 2 ~ \mathbb{E}
\Bigg\{  
\bigg( 
\sum^{b}_{u=a} h_{\rm{I}}[k,u] s^{(n)}_{{\rm{I}}}[k-u] - h_{\rm{Q}}[k,u] s^{(n)}_{{\rm{Q}}}[k-u]
\bigg)^2
\\& 
~~~~~~~~~~~~
\bigg( 
\sum^{b}_{u=a} h_{\rm{I}}[k,u] s^{(n)}_{{\rm{Q}}}[k-u] + h_{\rm{Q}}[k,u] s^{(n)}_{{\rm{I}}}[k-u]
\bigg)^2
\Bigg\}
\end{split}
\end{equation}

\noindent where (i) comes from the fact that $-h_{\rm{Q}}[k,u]$ and $h_{\rm{Q}}[k,u]$ have the same distribution, i.e. both are Gaussian random variables with zero mean and variance of $\displaystyle\frac{\sigma^2_{h_u}}{2}$. After some extensive mathematical manipulations and using the facts that (i) in-phase and quadrature components are independent (ii) channel taps with different delays are independent, one can show that

\begin{equation} \label{eq: indepency 2}
\begin{split} 
&\mathbb{E}
\Bigg\{  
\bigg( 
\sum^{b}_{u=a} h_{\rm{I}}[k,u] s^{(n)}_{{\rm{I}}}[k-u] - h_{\rm{Q}}[k,u] s^{(n)}_{{\rm{Q}}}[k-u]
\bigg)^4
\Bigg\}
\\&= 
2~\mathbb{E}
\big\{  
 s^{(n)}_{{\rm{I}}}[k-u]^4
\big\}  \sum^{b}_{u=a} \mathbb{E}
\big\{ h_{\rm{I}}[k,u]^4 \big\} ~~+
\\& 
~~~~ 12~ \mathbb{E}
\big\{  
 s^{(n)}_{{\rm{I}}}[k-u]^2
\big\}^2 \sum_{r = a}^{r = b} \sum_{l = a, l \neq r}^{l = b} \mathbb{E}
\big\{ h_{\rm{I}}[k,l]^2 \big\} \mathbb{E}
\big\{ h_{\rm{I}}[k,r]^2 \big\} 
\\& 
~~~~ + 6~ \mathbb{E}
\big\{  
 s^{(n)}_{{\rm{I}}}[k-u]^2
\big\}^2 \bigg(   \mathbb{E}
\big\{ h_{\rm{I}}[k,u]^2 \big\} \bigg)^2
\\&= 
 \frac{9}{8}\sigma^4_{s} \sum^{b}_{u=a} \sigma^4_{h_u} 
+ \frac{3}{4}~ \sigma^4_{s}   \sum_{r = a}^{r = b} \sum_{l = a, l \neq r}^{l = b} \sigma^2_{h_l} \sigma^2_{h_r} 
+
\frac{3}{8}~ \sigma^4_{s} \bigg( \sum_{l = a}^{b} \sigma^2_{h_l} \bigg)^2
\end{split}
\end{equation}

Similarly, one can verify that

\begin{equation} \label{eq: indepency 3}
\begin{split} 
&\mathbb{E}
\Bigg\{  
\bigg( 
\sum^{b}_{u=a} h_{\rm{I}}[k,u] s^{(n)}_{{\rm{I}}}[k-u] - h_{\rm{Q}}[k,u] s^{(n)}_{{\rm{Q}}}[k-u]
\bigg)^2 
\\& 
\bigg( 
\sum^{b}_{u=a} h_{\rm{I}}[k,u] s^{(n)}_{{\rm{Q}}}[k-u] + h_{\rm{Q}}[k,u] s^{(n)}_{{\rm{I}}}[k-u]
\bigg)^2
\Bigg\}
\\&= 
 4~ \mathbb{E}
\big\{  
 s^{(n)}_{{\rm{I}}}[k-u]^2
\big\}^2 \sum_{l = a}^{l = b} \mathbb{E}
\big\{ h_{\rm{I}}[k,l]^4 \big\} ~~+
\\& 
16~ \mathbb{E}
\big\{  
 s^{(n)}_{{\rm{I}}}[k-u]^2
\big\}^2 
\sum_{r = a}^{r = b} \sum_{l = a, l \neq r}^{l = b} \mathbb{E}
\big\{ h_{\rm{I}}[k,l]^2 \big\} \mathbb{E}
\big\{ h_{\rm{I}}[k,r]^2 \big\} 
\\& 
+
\bigg( 4~ \mathbb{E}
\big\{  
 s^{(n)}_{{\rm{I}}}[k-u]^4
\big\} - 8~ \mathbb{E}
\big\{  
 s^{(n)}_{{\rm{I}}}[k-u]^2
\big\}^2 \bigg)  
\\& 
~~~~~~~~~~ \times
\sum_{l = a}^{l = b} \big( \mathbb{E}
\big\{ h_{\rm{I}}[k,l]^2 \big\} \big)^2
\\&= 
\sigma^4_{s} \bigg( \sum^{b}_{u=a} \sigma^4_{h_u} +   \sum_{r = a}^{r = b} \sum_{l = a, l \neq r}^{l = b} \sigma^2_{h_l} \sigma^2_{h_r} \bigg)
\end{split}
\end{equation}

Substituting Equations \eqref{eq: indepency 2} and \eqref{eq: indepency 3} in Equation \eqref{eq: indepency 1} and then in Equation \eqref{eq: indepency 0} yields Equation \eqref{eq : weights}. One can easily show that the variance of the \ac{som} of a complex Gaussian random variable $w$ with zero mean and variance $\sigma^2_n$ is 
\begin{equation}
    \sigma^2_{\vert w \vert^2} = 2 \sigma^2_n.
\end{equation}
\noindent This completes the proof. 
\end{proof}

\IEEEpeerreviewmaketitle

\bibliographystyle{IEEEtran}
\bibliography{IEEEabrv,Reference}

\end{document}